\newcommand{\goesto}{\rightarrow}

\newcommand{\VanD}{\Delta\!^{1/2}}

\newcommand{\symmeq}{\stackrel{.}{=}}

\documentclass[prd,aps,eqsecnum,showpacs]{revtex4}
\usepackage{amsmath}
\usepackage{amssymb}
\begin{document}

\title{Noise Kernel and Stress Energy Bi-Tensor of Quantum Fields in
Hot Flat Space and Gaussian Approximation in the Optical
Schwarzschild Metric}

\author{Nicholas G. Phillips}
\email{Nicholas.G.Phillips@gsfc.nasa.gov}
\affiliation{ SSAI, Laboratory for Astronomy and Solar Physics, Code 685, 
             NASA/GSFC, Greenbelt, Maryland 20771}

\author{B. L. Hu}
\email{hub@physics.umd.edu}
\affiliation{Department of Physics, University of Maryland, College
             Park, Maryland 20742-4111}

\date{umdpp 03-016, Submitted to Phys. Rev. D, September 17, 2002}

\begin{abstract}
Continuing our investigation of the regularization of the noise
kernel in curved spacetimes [N. G.  Phillips and  B. L. Hu, Phys.
Rev. D {\bf 63},  104001 (2001)] we adopt the modified point
separation scheme for the class of optical spacetimes using the
Gaussian approximation for the Green functions a la
Bekenstein-Parker-Page. In the first example we derive the
regularized noise kernel for a thermal field in flat space. It is
useful for black hole nucleation considerations. In the second
example of an optical Schwarzschild spacetime we obtain a finite
expression for the noise kernel at the horizon and recover the
hot flat space result at infinity. Knowledge of the noise kernel
is essential for studying issues related to black hole horizon
fluctuations and Hawking radiation backreaction. We show that the
Gaussian approximated Green function which works surprisingly well
for the stress tensor at the Schwarzschild horizon produces
significant error in the noise kernel there. We identify the
failure as occurring at the fourth covariant derivative order.
\end{abstract}

\maketitle

%%%%%%%%%%%%%LATEX %%%%%%%%%%%%%%%%%%%%

%\section{Introduction}

%%%%%%%%%%%%%%%%

\section{Introduction}
\label{sec-ch4-intro}

The noise kernel is the vacuum expectation value of the
stress-energy bi-tensor for a quantum field. In curved spacetime
field theory \cite{BirDav,Full89,Wald94,Grib} it plays a  role in
stochastic semiclassical gravity
\cite{Physica,ELE,CamHu,stogra,MarVer,HVErice} similar to the
expectation value of the stress-energy tensor in semiclassical
gravity \cite{scg}. The noise kernel being a measure of the
fluctuations of the stress tensor of quantum fields enters in a
great variety of issues and problems ranging from the validity of
semiclassical gravity to spacetime foams, from structure
formation in the early universe to fluctuations of the black hole
horizon and trans-Planckian physics (for a review of this subject
see \cite{HVErice} ). Noise kernel in hot flat space, one of the
two examples considered here, has been studied by Campos and Hu
\cite{CamHu}. It is useful for performing a nonequilibrium quantum
field theoretical analysis of the black hole nucleation problems
beyond Euclidean thermodynamics \cite{GPY,WhiYor}.

In Paper I \cite{PH1}, we have derived a general expression for
the noise kernel of a quantum scalar field in an arbitrary curved
spacetime as products of covariant derivatives of the quantum
field's Green function. It is finite when the noise kernel is
evaluated for distinct pairs of points (and non-null points for a
massless field). We also showed explicitly the trace of the noise
kernel vanishes, confirming there is no noise associated with the
trace anomaly. This holds regardless of issues of regularization
of the noise kernel.

The noise kernel as a two point function of the stress energy
tensor diverges  as the pair of points are brought together,
sharing the ubiquitous ultraviolet divergence present in ordinary
(point-defined) quantum field theory. To calculate a regularized
noise kernel, similar to what was done before for the simpler
stress energy tensor, it is desirable to have at the start an
analytic and finite regularized expression for the Green
function. When one can carry out a mode decomposition of the
invariant operator and find an analytic solution to the mode
functions there are established ways to proceed. % regularize the ultraviolet divergences.
For such cases the quantum stress tensor and its fluctuations can
be determined using some regularization scheme, from the simple
normal ordering \cite{KuoFor,WuFor} or smeared field \cite{PH0} in
Minkowski and Casimir spaces to the elegant $\zeta$-function or
the powerful point-separation methods, as demonstrated for the
Einstein Universe or for the Casimir effect by the authors
earlier \cite{PH97,PH0} and others
\cite{Elizalde,Kirsten,Camporesi}. Unfortunately, there are many
important geometries for which an exact analytic expression of the
mode functions is not available, such as the Schwarzschild black
hole spacetime.

In this series of papers we work with the covariant point
separation regularization method
\cite{DeWitt65,Christensen,ChristWinnipeg} in its modified form
\cite{Adler,Wald} to derive a finite expression for the
coincident limit of the noise kernel. The expression derived
in Paper I \cite{PH1}
for the noise kernel of a scalar field is completely general and
can be used with or without consideration of the renormalization
of the Green function. Also, the result there for the coincident
limit
holds for all choices of the Green function and the metric
provided that the Green function has a meaningful coincident
limit. In this paper and the next one, we apply this formal
procedure to specific spacetimes of physical interest. We do this
by working with an analytic form of the Green function.  When
such a form is available one can carry out an  end point expansion
displaying  the ultraviolet divergence. Subtraction of the
Hadamard ansatz
expressed as a series expansion
will render this Green function finite in the coincident limit.
With this, one can calculate the noise kernel for a variety of
spacetimes.

An analytic form is obtained by invoking the Gaussian
approximation introduced by Bekenstein and Parker
\cite{BekPark81}. For a massless scalar field in an ultra-static
spacetime whose metric has an optical form (one where the
Euclidean time $\tau$-time component of the metric $g_{\tau\tau}
= 1$ ) this provides a closed expression for the Green function.
In this paper we use the Gaussian approximation for the Green
function for such quantum fields to evaluate the noise kernel in
the following optical metrics : hot flat space, and the optical
Schwarzschild spacetime, which is conformal to the Schwarzschild
metric.  For hot flat space, the Gaussian Green function is
exact. For optical Schwarzschild, the Gaussian Green function is
known to be a fairly good approximation for calculating the
stress tensor \cite{Page82}, which involves second covariant
derivatives of the Green function. We will carry out this
calculation for the noise kernel which requires up to four
covariant derivatives of the Green function. Thus the validity of
the Gaussian approximation will be tested to its new limit. A
reliable check is provided by the trace of the noise kernel, which
for massless conformal fields should be zero.

Thus the goal of this paper is threefold: First,  to present the
detailed steps in the calculation of the regularized noise kernel
for a quantum scalar field in a  general curved spacetime using
the modified point-separation scheme. Second, to derive the
regularized noise kernel for a thermal field in flat space.
Third, to determine the range of validity of the Gaussian
(Bekenstein-Parker-Page) approximation to the Green function by
examining the error in the noise kernel expression for the optical
Schwarzschild spacetime.

We present an outline of the calculation as follows: In Section
\ref{sec-ch4-gauss} we give a brief description of the Gaussian
approximation to the Green function \cite{BekPark81} for
ultrastatic spacetimes \cite{Page82}.  In Section
\ref{sec-ch4-optnoise} we consider the regularization of the heat
kernel in the class of optical metrics.  We expand this Green
function in an end point series so that it can be separated into
a divergent piece and a finite remainder. The divergent term is
independent of the approximation since we know this structure
must be of a general form given by the Hadamard ansatz.
After this is subtracted, we have a series expansion of the
renormalized Green function. We then substitute this expansion in
the general expression obtained in Paper I
for the  coincident limit of the noise kernel. The resulting
expression is quite long and formal.  At this point one can
introduce the specific metric of interest and determine the
component values of the Green function expansion tensors (by
symbolic computation).
From this we can readily generate all the needed component values
of the coincident limits of the covariant derivatives of the
Green function, along with the covariant derivatives of the
coincident limits. These explicitly evaluated tensors are then
substituted in the general expression obtained in Paper I
to get the final result. \footnote{We must stress, as we did in
Paper I, that though all these are done on a computer, no
numerical approximation is used. All work is done symbolically in
terms of the explicit functional form of the metric and the
parameters of the field. The final results are exact to the
extent that the analytic form of the Green function is exact.} We
give two examples in Section
\ref{sec-Examples}. For the case of hot flat space
we derive the variance of the energy and pressure density for a
quantum field at finite temperature. This is a useful compendium
to the results obtained earlier \cite{PH0} for quantum fields in
Minkowski and Casimir geometries in reference to issues like the
validity of semiclassical gravity \cite{KuoFor,PH0,HP0,ForWu}.
For  a massless, conformally coupled field in the optical
Schwarzschild (the ultrastatic spacetime conformal to the
Schwarzschild black hole), we obtain for the regularized noise
kernel at spatial infinity the same result as a thermal field in
flat space, as it should, and a finite result at the horizon in a
state conformally-related to the Hartle-Hawking state. However the
latter expression computed with the Gaussian approximation  has a
non-vanishing trace. In Section \ref{sec-errGauss} we study the
nature and source of this error by examining the validity of the
Gaussian approximation at successive orders. It works reasonably
well to the third covariant derivative order. The inadequacy of
the Gaussian approximation to the Green function for the
calculation of the noise kernel arises from the Green function's
failure to satisfy the field equation of the scalar field at the
fourth order.

\section{Gaussian Approximation}
\label{sec-ch4-gauss}

We give a brief description of the Gaussian approximation to the
Green function for quantum fields in optical spacetimes a la
Bekenstein-Parker\cite{BekPark81} and Page \cite{Page82}.

In the Schwinger-DeWitt proper-time formalism \cite{SchDeW} the
Green function is expressed in terms of the heat kernel
$K(x,y,s)$ via
\begin{equation}
G(x,y) 
  = 
   \int_0^\infty 
     K(x,y,s) 
       ds.
\label{Chap4-propertime}
\end{equation}
where the heat kernel satisfies
\begin{equation} 
\left[
   \frac{\partial}{\partial s} - \left ( \square -\frac{R}{6} \right )
\right] K(x,y,s)  = 0, \quad
 K(x,y,0) = \delta(x-y)
\end{equation} The optical metric for an ultrastatic spacetime has the
product form \begin{equation} ds^2 = g_{ab}dx^a dx^b = d\tau^2 + g_{ij}dx^i
dx^j \end{equation} We assume in the Euclidean sector  the imaginary time
dimension is periodic with period
$2\pi/\kappa=1/T$ with $T$ the temperature.
For a black hole,  $\kappa$ is the surface
gravity but can be regarded as a temperature parameter here. This
form of the metric allows the kernel to take on the product form
\begin{equation}
 K(x,y,s) =  K_1(\tau,\tau',s)  K_3({\bf x},{\bf y},s)
\label{Chap4-Kproduct}
\end{equation}
with each of the kernels satisfying
\begin{subequations}\begin{eqnarray}
\left(
   \frac{\partial}{\partial s} - \frac{\partial^2}{\partial \tau^2}
\right) K_1(\tau,\tau',s)  &=& 0,
\label{Chap4-K1eqn}\\
\left[
   \frac{\partial}{\partial s} - \left(\nabla_i\nabla^i -\frac{R}{6}\right)
\right] K_3({\bf x},{\bf y},s)  &=& 0 \label{Chap4-K3eqn}
\end{eqnarray}\end{subequations}

Equation (\ref{Chap4-K1eqn}) has the periodic solution 
\begin{equation}
  K_1(\tau,\tau',s) = \frac{\kappa}{2\pi}\sum_{n=-\infty}^{\infty}
\exp\left(-\kappa^2 n^2 s + i\kappa n\Delta\tau \right)
\label{chap4-K1soln} 
\end{equation} 
($\Delta\tau = \tau-\tau'$). Equation
(\ref{Chap4-K3eqn}) in general is difficult to solve, but
Bekenstein and Parker\cite{BekPark81} find an approximate solution using the
Gaussian approximation to the path integral representation. For
$K_3$, it takes the form \begin{equation} K_{3{\rm Gauss}}({\bf x},{\bf y},s)
= \frac{{}^{(3)}\!\VanD}{\left(4\pi s\right)^\frac{3}{2}}
   \exp\left( -\frac{{}^{(3)}\!\sigma}{2s} \right)
\label{chap4-K3soln} \end{equation} where ${}^{(3)}\!\sigma$ is the world
function for the three-dimensional spatial geometry. In the above
${}^{(3)}\!\VanD$ is the VanVleck-Morette determinant for the spatial
geometry.
Since we have an optical metric, the four-dimensional world
function is $\sigma = {}^{(3)}\!\sigma +\Delta\tau^2/2$ and there is no
difference between the three and four dimensional $\VanD$.

Since the complete Hadamard-Minakshisundaram-Pleijel-DeWitt
expansion \cite{DeWitt65} would be \begin{equation} K_{3} = K_{3{\rm
Gauss}}\sum_{n=0}^\infty a_n({\bf x},{\bf y}) s^n
\label{Chap4-SQW-series} \end{equation} the Gaussian approximation is
equivalent to only taking the first term in this power series.

By putting (\ref{chap4-K1soln}) and (\ref{chap4-K3soln}) back into
(\ref{Chap4-Kproduct}) and carrying out the integration
(\ref{Chap4-propertime}), Page obtains
\begin{equation} G_{\rm Gauss}(x,y) =
\frac{\kappa\VanD}{8\pi^2 r}
  \frac{\sinh\kappa r}{\left(\cosh\kappa r-\cos\kappa\tau\right)}
\label{Chap4-GGauss}
\end{equation}
as the Gaussian approximation for the Green function,
where $r=\left(2\,{}^{(3)}\!\sigma\right)^\frac{1}{2}$.

%======================================================================
%======================================================================
\section{Noise Kernel in Optical Spacetimes}
\label{sec-ch4-optnoise}

In this section, the noise kernel for an ultrastatic spacetime
with an optical metric is determined. For this class of
geometries, we start directly with (\ref{Chap4-GGauss}) for
the Green function. The first step is to expand this Green
function about the coincident limit. Since the noise kernel has
terms with at most four covariant derivatives, this expansion
needs to be to fourth order in $\sigma^a$, or second order in
$\sigma=\sigma^p\sigma_p/2$, and fourth order in $\Delta\tau$.
Doing this expansion yields
 \begin{eqnarray} G_{\rm
Gauss } &=& {\frac{{\Delta^{{\frac{1}{2}}}}}{8\,{{\pi
}^2}\,\sigma}} + {\frac{{\Delta^{{\frac{1}{2}}}}}{8\,{{\pi }^2}}}
\left\{ {\frac{{{\kappa}^2}}{6}}
  + {\frac{{{\kappa}^4}}{180}} \left( 2\,{{{\Delta\tau}}^2} - \sigma \right)
\right. \cr && \left. \hspace{20mm}
+ {\frac{{{\kappa}^6}}{3780}} \left( 4\,{{{\Delta\tau}}^4} -
6\,{{{\Delta\tau}}^2}\,\sigma + {{\sigma}^2} \right)
\right\}
+O\left(\sigma^\frac{5}{2},\delta\tau^5\right)
\label{Opt-Gseries1}
\end{eqnarray}
By subtracting from the Gaussian Green function the Hadamard ansatz
\begin{equation}
S(x,y) =
{\frac{1}{16\,{{\pi }^2}}} \left( {\frac{2\,{\Delta^{{\frac{1}{2}}}}}{\sigma}}
+ \sigma\,w_{1} +
  {{\sigma}^2}\,w_{2} \right)
 + O\left(\sigma^3\right)
\end{equation} (the $V(x,x')$ term is absent since there is no $\log\sigma$
divergence present in the expansion of the Gaussian approximation
to the Green function) we get the renormalized Green function
\begin{equation} G_{\rm ren} = G_{\rm gauss} - S. \label{Opt-Gren1} \end{equation} The
divergent term present (\ref{Opt-Gseries1}) is cancelled by the
divergent term from the Hadamard ansatz.

Since our main interest here is to determine the coincident limit
of the noise kernel, we next turn to developing the
series expansion \begin{eqnarray} G_{\rm ren} &=& \frac{1}{(4\pi)^2} \left(
    G_{{\rm ren}}^{(0)} +
    \sigma{}^{;}{}^{p}\,G_{{\rm ren}}^{(1)}{}_{p} +
  \sigma{}^{;}{}^{p}\,\sigma{}^{;}{}^{q}\,G_{{\rm ren}}^{(2)}{}_{p}{}_{q} +
  \sigma{}^{;}{}^{p}\,\sigma{}^{;}{}^{q}\,\sigma{}^{;}{}^{r}\,
   G_{{\rm ren}}^{(3)}{}_{p}{}_{q}{}_{r} +
\right. \cr && \hspace{12mm} \left. +
    \sigma{}^{;}{}^{p}\,\sigma{}^{;}{}^{q}\,\sigma{}^{;}{}^{r}\,
  \sigma{}^{;}{}^{s}\,G_{{\rm ren}}^{(4)}{}_{p}{}_{q}{}_{r}{}_{s}
\right) \label{Opt-Gren2} \end{eqnarray} of the regularized Green
function. With this, it will be straightforward to compute
the coincident limits of the various covariant derivatives needed.
We start by assuming the expansions
\begin{subequations}\begin{eqnarray}
{\Delta^{{\frac{1}{2}}}} &\approx& 1 +
\sigma{}^{;}{}^{p}\,\sigma{}^{;}{}^{q}\,\Delta^{(2)}{}_{p}{}_{q} +
  \sigma{}^{;}{}^{p}\,\sigma{}^{;}{}^{q}\,\sigma{}^{;}{}^{r}\,
   \Delta^{(3)}{}_{p}{}_{q}{}_{r} +
  \sigma{}^{;}{}^{p}\,\sigma{}^{;}{}^{q}\,\sigma{}^{;}{}^{r}\,
   \sigma{}^{;}{}^{s}\,\Delta^{(4)}{}_{p}{}_{q}{}_{r}{}_{s} \\
{{{\Delta\tau}}^2} &\approx&
\sigma{}^{;}{}^{p}\,\sigma{}^{;}{}^{q}\,\delta\tau^{(2)}{}_{p}{}_{q} +
  \sigma{}^{;}{}^{p}\,\sigma{}^{;}{}^{q}\,\sigma{}^{;}{}^{r}\,
   \delta\tau^{(3)}{}_{p}{}_{q}{}_{r} +
  \sigma{}^{;}{}^{p}\,\sigma{}^{;}{}^{q}\,\sigma{}^{;}{}^{r}\,
   \sigma{}^{;}{}^{s}\,\delta\tau^{(4)}{}_{p}{}_{q}{}_{r}{}_{s} \\
w_{1} &\approx& w^{(0)}_{1} + \sigma{}^{;}{}^{p}\,w^{(1)}_{1}{}_{p} +
  \sigma{}^{;}{}^{p}\,\sigma{}^{;}{}^{q}\,w^{(2)}_{1}{}_{p}{}_{q} \\
w_{2} &\approx& w^{(0)}_{2}
\end{eqnarray} \label{Opt-expands} \end{subequations}
The
specific values of the expansion tensors in these series are
derived in the Appendices. Carrying out the subtraction
(\ref{Opt-Gren1}) and substituting the expansions
(\ref{Opt-expands}), we find the expansions tensors in
(\ref{Opt-Gren2}) are 
\begin{subequations}\begin{eqnarray}
G_{{\rm ren}}^{(0)}                 &=& {\frac{{{\kappa}^2}}{3}} \\
G_{{\rm ren}}^{(1)}{}_{a}                 &=& 0 \\
G_{{\rm ren}}^{(2)}{}_{a}{}_{b}         &=&
 {\frac{{{\kappa}^2}}{3}} \Delta^{(2)}{}_{a}{}_{b}
+{\frac{{{\kappa}^4}}{180}} \left( 4\,\delta\tau^{(2)}{}_{a}{}_{b} -
g{}_{a}{}_{b} \right)
-\frac{1}{2} w^{(0)}_{1}\,g{}_{a}{}_{b}
\\
G_{{\rm ren}}^{(3)}{}_{a}{}_{b}{}_{c}         &=&
 {\frac{{{\kappa}^2}}{3}} \Delta^{(3)}{}_{a}{}_{b}{}_{c}
+{\frac{{{\kappa}^4}}{45}} \delta\tau^{(3)}{}_{a}{}_{b}{}_{c}
-\frac{1}{2} g{}_{a}{}_{b}\,w^{(1)}_{1}{}_{c}
\\
G_{{\rm ren}}^{(4)}{}_{a}{}_{b}{}_{c}{}_{d}        &=&
 {\frac{{{\kappa}^2}}{3}} \Delta^{(4)}{}_{a}{}_{b}{}_{c}{}_{d}
+{\frac{{{\kappa}^4}}{180}} \left(
4\,\Delta^{(2)}{}_{a}{}_{b}\,\delta\tau^{(2)}{}_{c}{}_{d} +
  4\,\delta\tau^{(4)}{}_{a}{}_{b}{}_{c}{}_{d} -
  \Delta^{(2)}{}_{c}{}_{d}\,g{}_{a}{}_{b} \right)
\cr &&
+{\frac{{{\kappa}^6}}{7560}} \left(
16\,\delta\tau^{(2)}{}_{a}{}_{b}\,\delta\tau^{(2)}{}_{c}{}_{d} -
  12\,\delta\tau^{(2)}{}_{c}{}_{d}\,g{}_{a}{}_{b} +
  g{}_{a}{}_{b}\,g{}_{c}{}_{d} \right)
\cr &&
-\frac{1}{4} g{}_{a}{}_{b}\,\left( w^{(0)}_{2}\,g{}_{c}{}_{d} +
    2\,w^{(2)}_{1}{}_{c}{}_{d} \right)
\end{eqnarray}\end{subequations}

Using the explicit forms of the expansion tensor values,
(\ref{Apx-D2rule}), (\ref{Apx-D3rule}), (\ref{Apx-D4rule}) and
(\ref{Apx-dtrules}), we get
\begin{subequations}\begin{eqnarray}
G_{{\rm ren}}^{(2)}{}_{a}{}_{b}        &=&
 {\frac{{{\kappa}^2}}{36}} R{}_{a}{}_{b}
+{\frac{{{\kappa}^4}}{180}} \left(
4\,{\delta}{}_{a}{}^{\tau}\,{\delta}{}_{b}{}^{\tau} - g{}_{a}{}_{b} \right)
\\
G_{{\rm ren}}^{(3)}{}_{a}{}_{b}{}_{c}     &=&
-{\frac{{{\kappa}^2}}{72}} R{}_{a}{}_{b}{}_{;}{}_{c}
+{\frac{{{\kappa}^4}}{45}} \Gamma{}^{\tau}{}_{a}{}_{b}\,{\delta}{}_{c}{}^{\tau}
\\
G_{{\rm ren}}^{(4)}{}_{a}{}_{b}{}_{c}{}_{d}  &=&
 {\frac{{{\kappa}^2}}{4320}} \left( 18\,R{}_{a}{}_{b}{}_{;}{}_{c}{}_{d} +
5\,R{}_{a}{}_{b}\,R{}_{c}{}_{d} +
  4\,R{}_{p}{}_{a}{}_{q}{}_{b}\,R{}_{c}{}^{p}{}_{d}{}^{q} \right)
\cr &&
+{\frac{{{\kappa}^4}}{2160}} \left(
12\,\Gamma{}^{\tau}{}_{a}{}_{b}\,\Gamma{}^{\tau}{}_{c}{}_{d} -
  16\,\Gamma{}^{\tau}{}_{a}{}_{b}{}_{;}{}_{c}\,{\delta}{}_{d}{}^{\tau} +
  4\,{\delta}{}_{a}{}^{\tau}\,{\delta}{}_{b}{}^{\tau}\,R{}_{c}{}_{d} -
  g{}_{a}{}_{b}\,R{}_{c}{}_{d} \right)
\cr &&
+{\frac{{{\kappa}^6}}{7560}} \left(
16\,{\delta}{}_{a}{}^{\tau}\,{\delta}{}_{b}{}^{\tau}\,
   {\delta}{}_{c}{}^{\tau}\,{\delta}{}_{d}{}^{\tau} -
  12\,{\delta}{}_{a}{}^{\tau}\,{\delta}{}_{b}{}^{\tau}\,g{}_{c}{}_{d} +
  g{}_{a}{}_{b}\,g{}_{c}{}_{d} \right)
\cr &&
-\frac{1}{4} g{}_{a}{}_{b}\,\left( w^{(0)}_{2}\,g{}_{c}{}_{d} +
    2\,w^{(2)}_{1}{}_{c}{}_{d} \right)
\label{Opt-Gvalues2} 
\end{eqnarray}\end{subequations}
 where we have also used 
\begin{equation}
w^{(0)}_{1} = -\frac{1}{240} \left( R{}_{;}{}_{p}{}^{p} -
         R{}_{p}{}_{q}\,R{}^{p}{}^{q} +
        R{}_{p}{}_{q}{}_{r}{}_{s}\,R{}^{p}{}^{q}{}^{r}{}^{s}\right) = 0,
\end{equation}
along with $w^{(1)}_{1}{}_{a} = 0$, which hold for ultrastatic metrics.

For flat space, these tensors reduce to
\begin{subequations}\begin{eqnarray}
G_{{\rm ren}}^{(2)}{}_{a}{}_{b}        &=&
{\frac{{{\kappa}^4}}{180}} \left( -{\eta}{}_{a}{}_{b} +
4\,{\eta}{}_{a}{}^{\tau}\,{\eta}{}_{b}{}^{\tau} \right)
\\
G_{{\rm ren}}^{(3)}{}_{a}{}_{b}{}_{c}        &=& 0
\\
G_{{\rm ren}}^{(4)}{}_{a}{}_{b}{}_{c}{}_{d}        &=&
{\frac{{{\kappa}^6}}{7560}} \left( {\eta}{}_{a}{}_{b}\,{\eta}{}_{c}{}_{d} -
  12\,{\eta}{}_{a}{}^{\tau}\,{\eta}{}_{b}{}^{\tau}\,{\eta}{}_{c}{}_{d} +
  16\,{\eta}{}_{a}{}^{\tau}\,{\eta}{}_{b}{}^{\tau}\,{\eta}{}_{c}{}^{\tau}\,
   {\eta}{}_{d}{}^{\tau} \right)
\label{Opt-flat-co}
\end{eqnarray}\end{subequations}

Now that we know the end point series expansion (\ref{Opt-Gren2})
of $G_{{\rm ren}}$, the coincident limit of terms with up to four
covariant derivatives are computed. We simply differentiate
the series (\ref{Opt-Gren2}) and then use the results from
Appendix (\ref{apx-sigma}) for the coincident limits of the
covariant derivatives of the world function $\sigma$. The results
are \begin{subequations}\begin{eqnarray}
16\,{{\pi }^2}\,{\left[G_{{\rm ren}}\right]} &=&  G_{{\rm ren}}^{(0)}
\\
16\,{{\pi }^2}\,{\left[G_{{\rm ren}}{}_{;}{}_{a}\right]} &=& G_{{\rm
ren}}^{(0)}{}_{;}{}_{a}
\\
16\,{{\pi }^2}\,{\left[G_{{\rm ren}}{}_{;}{}_{a}{}_{b}\right]} &=& G_{{\rm
ren}}^{(0)}{}_{;}{}_{(}\!{}_{a}{}_{b}\!{}_{)} +
  2\,G_{{\rm ren}}^{(2)}{}_{(}\!{}_{a}{}_{b}\!{}_{)}
\\
16\,{{\pi }^2}\,{\left[G_{{\rm ren}}{}_{;}{}_{a}{}_{b}{}_{c}\right]} &=&
 6\,\left( G_{{\rm ren}}^{(2)}{}_{(}\!{}_{a}{}_{b}{}_{;}{}_{c}\!{}_{)} +
    G_{{\rm ren}}^{(3)}{}_{(}\!{}_{a}{}_{b}{}_{c}\!{}_{)} \right)  + G_{{\rm
ren}}^{(0)}{}_{;}{}_{a}{}_{b}{}_{c}
\\
16\,{{\pi }^2}\,{\left[G_{{\rm ren}}{}_{;}{}_{a}{}_{b}{}_{c}{}_{d}\right]} &=&
  12\,\left( 2\,G_{{\rm ren}}^{(3)}{}_{(}\!{}_{a}{}_{b}{}_{c}{}_{;}
      {}_{d}\!{}_{)} + G_{{\rm ren}}^{(2)}{}_{(}\!{}_{a}{}_{b}{}_{;}{}_{c}
     {}_{d}\!{}_{)} + 2\,G_{{\rm ren}}^{(4)}{}_{(}\!{}_{a}{}_{b}{}_{c}
      {}_{d}\!{}_{)} \right)  \cr
&&+\frac{2}{3} \left(
  G_{{\rm ren}}^{(2)}{}_{p}{}_{a}\,
  \left( R{}_{b}{}_{c}{}_{d}{}^{p} - 2\,R{}_{b}{}_{d}{}_{c}{}^{p} \right)  +
G_{{\rm ren}}^{(2)}{}_{p}{}_{b}\,
  \left( R{}_{a}{}_{c}{}_{d}{}^{p} - 2\,R{}_{a}{}_{d}{}_{c}{}^{p} \right)
\right. \cr && \hspace{4mm} + \left.
  G_{{\rm ren}}^{(2)}{}_{p}{}_{c}\,
  \left( R{}_{a}{}_{b}{}_{d}{}^{p} - 2\,R{}_{a}{}_{d}{}_{b}{}^{p} \right)  +
G_{{\rm ren}}^{(2)}{}_{p}{}_{d}\,
  \left( R{}_{a}{}_{b}{}_{c}{}^{p} - 2\,R{}_{a}{}_{c}{}_{b}{}^{p} \right)
\right) \cr
&&+ G_{{\rm ren}}^{(0)}{}_{;}{}_{a}{}_{b}{}_{c}{}_{d}
\label{Opt-CDGr-to-Grn}
\end{eqnarray}\end{subequations}

We now have all the information we need to compute the coincident
limit of the noise kernel (see Eqn (3.24) of Paper I). Since the
point separated noise kernel $N_{abc'd'}(x,y)$ involves covariant
derivatives at the two points at which it has support, when we
take the coincident limit we can use Synge's theorem to move the
derivatives acting at the second point $y$ to ones acting at the
first point $x$. Due to the length of the expression for the noise
kernel, we will here give an example of the calculation by
examining a single term. The complete expression for the
coincident limit of the point separated noise kernel can be found
in Paper I as Eqn (4.16).

Consider a typical term from the noise kernel functional
(Eqn (3.24) of Paper I):
\begin{equation}
G_{{\rm ren}}{}_{;}{}_{c'}{}_{b}\,G_{{\rm ren}}{}_{;}{}_{d'}{}_{a} + G_{{\rm
ren}}{}_{;}{}_{c'}{}_{a}\,G_{{\rm ren}}{}_{;}{}_{d'}{}_{b}
\end{equation}
As was derived in Paper I,
the noise kernel itself is related to the noise kernel functional via
\begin{equation}
N_{abc'd'} = N_{abc'd'}\left[G_{ren}(x,y)\right]
                + N_{abc'd'}\left[G_{ren}(y,x)\right].
\end{equation}
We account for this by adding the
same term but with the roles of $x$ and $y$ reversed.
Taken together, we need to analyze
\begin{equation} G_{{\rm ren}}{}_{;}{}_{c'}{}_{b}\,G_{{\rm
ren}}{}_{;}{}_{d'}{}_{a} + G_{{\rm
ren}}{}_{;}{}_{a'}{}_{d}\,G_{{\rm ren}}{}_{;}{}_{b'}{}_{c} +
G_{{\rm ren}}{}_{;}{}_{c'}{}_{a}\,G_{{\rm
ren}}{}_{;}{}_{d'}{}_{b} + G_{{\rm
ren}}{}_{;}{}_{a'}{}_{c}\,G_{{\rm ren}}{}_{;}{}_{b'}{}_{d}, \end{equation}
in particular, its coincident limit:
\begin{equation} {\left[G_{{\rm
ren}}{}_{;}{}_{c'}{}_{b}\right]}\,
  {\left[G_{{\rm ren}}{}_{;}{}_{d'}{}_{a}\right]} + {\left[G_{{\rm
ren}}{}_{;}{}_{a'}{}_{d}\right]}\,
  {\left[G_{{\rm ren}}{}_{;}{}_{b'}{}_{c}\right]} + {\left[G_{{\rm
ren}}{}_{;}{}_{c'}{}_{a}\right]}\,
  {\left[G_{{\rm ren}}{}_{;}{}_{d'}{}_{b}\right]} + {\left[G_{{\rm
ren}}{}_{;}{}_{a'}{}_{c}\right]}\,
  {\left[G_{{\rm ren}}{}_{;}{}_{b'}{}_{d}\right]}
\end{equation}
We apply Synge's theorem to remove any explicit reference to the
point $y$,
\begin{eqnarray}
&&\hspace{3mm}
\left( {\left[G_{{\rm ren}}{}_{;}{}_{a}\right]}{}_{;}{}_{d} -
    {\left[G_{{\rm ren}}{}_{;}{}_{a}{}_{d}\right]} \right) \,
  \left( {\left[G_{{\rm ren}}{}_{;}{}_{b}\right]}{}_{;}{}_{c} -
    {\left[G_{{\rm ren}}{}_{;}{}_{b}{}_{c}\right]} \right)  \cr
&& + \left( {\left[G_{{\rm ren}}{}_{;}{}_{d}\right]}{}_{;}{}_{a} -
    {\left[G_{{\rm ren}}{}_{;}{}_{a}{}_{d}\right]} \right) \,
  \left( {\left[G_{{\rm ren}}{}_{;}{}_{c}\right]}{}_{;}{}_{b} -
    {\left[G_{{\rm ren}}{}_{;}{}_{b}{}_{c}\right]} \right)  \cr
&& + \left( {\left[G_{{\rm ren}}{}_{;}{}_{a}\right]}{}_{;}{}_{c} -
    {\left[G_{{\rm ren}}{}_{;}{}_{a}{}_{c}\right]} \right) \,
  \left( {\left[G_{{\rm ren}}{}_{;}{}_{b}\right]}{}_{;}{}_{d} -
    {\left[G_{{\rm ren}}{}_{;}{}_{b}{}_{d}\right]} \right)  \cr
&& + \left( {\left[G_{{\rm ren}}{}_{;}{}_{c}\right]}{}_{;}{}_{a} -
    {\left[G_{{\rm ren}}{}_{;}{}_{a}{}_{c}\right]} \right) \,
  \left( {\left[G_{{\rm ren}}{}_{;}{}_{d}\right]}{}_{;}{}_{b} -
    {\left[G_{{\rm ren}}{}_{;}{}_{b}{}_{d}\right]} \right)
\end{eqnarray}
and express the results in terms of the expansion tensors
(\ref{Opt-CDGr-to-Grn} ) and find
\begin{equation} 8\,\left( G_{{\rm
ren}}^{(2)}{}_{a}{}_{d}\,G_{{\rm ren}}^{(2)}{}_{b}{}_{c} +
    G_{{\rm ren}}^{(2)}{}_{a}{}_{c}\,G_{{\rm ren}}^{(2)}{}_{b}{}_{d} \right)
\end{equation}
Though this may look relatively compact, when we use the
results (\ref{Opt-Gvalues2}) to get the final form,
in terms of the local geometry,
this is no longer the case. Making these substitutions,
our pair of terms become
\begin{eqnarray}
&& {\frac{{{\kappa}^4}}{2592\,{{\pi }^2}}} \left( R{}_{a}{}_{d}\,R{}_{b}{}_{c}
+ R{}_{a}{}_{c}\,R{}_{b}{}_{d} \right)
\cr &&
+ {\frac{{{\kappa}^6}}{12960\,{{\pi }^2}}} \left\{
\left( 4\,{\delta}{}_{b}{}^{\tau}\,{\delta}{}_{d}{}^{\tau} -
     g{}_{b}{}_{d} \right) \,R{}_{a}{}_{c} +
  \left( 4\,{\delta}{}_{b}{}^{\tau}\,{\delta}{}_{c}{}^{\tau} -
     g{}_{b}{}_{c} \right) \,R{}_{a}{}_{d}
\right. \cr && \left. \hspace{5mm}
 + \left( 4\,{\delta}{}_{a}{}^{\tau}\,{\delta}{}_{d}{}^{\tau} -
     g{}_{a}{}_{d} \right) \,R{}_{b}{}_{c} +
  \left( 4\,{\delta}{}_{a}{}^{\tau}\,{\delta}{}_{c}{}^{\tau} -
     g{}_{a}{}_{c} \right) \,R{}_{b}{}_{d} \right\}
\cr &&
 + {\frac{{{\kappa}^8}}{64800\,{{\pi }^2}}} \left\{
32\,{\delta}{}_{a}{}^{\tau}\,{\delta}{}_{b}{}^{\tau}\,
  {\delta}{}_{c}{}^{\tau}\,{\delta}{}_{d}{}^{\tau}
+ g{}_{a}{}_{d}\,g{}_{b}{}_{c} + g{}_{a}{}_{c}\,g{}_{b}{}_{d}
\right. \cr && \left. \hspace{5mm}
-4\,\left( {\delta}{}_{b}{}^{\tau}\,{\delta}{}_{d}{}^{\tau}\,g{}_{a}{}_{c} +
    {\delta}{}_{b}{}^{\tau}\,{\delta}{}_{c}{}^{\tau}\,g{}_{a}{}_{d} +
    {\delta}{}_{a}{}^{\tau}\,{\delta}{}_{d}{}^{\tau}\,g{}_{b}{}_{c} +
    {\delta}{}_{a}{}^{\tau}\,{\delta}{}_{c}{}^{\tau}\,g{}_{b}{}_{d} \right)
\right\} \end{eqnarray} The noise kernel functional
%(\ref{PSChap-conformal-noise-kernel})
consists of 25 such terms, some quite a bit more involved. This is
especially true when a single Green function has four derivatives
acting on it. To gain insight into the physics we work with some
specific spacetime.\footnote{ Since this work for analyzing the
coincident limit of the noise kernel via the point separation
method is tailored to symbolic computation on the computer, our
method for deriving particular results is designed to take
maximum advantage of the computer.}

When we do choose a metric, we proceed by directly evaluating the
components of the expansion tensors (\ref{Opt-Gvalues2}). Once
these expansion tensor components are known, the actual
components of the coincident limits of the covariant derivatives
of $G_{{\rm ren}}$ are in turn computed using
(\ref{Opt-CDGr-to-Grn}). Then it is straightforward from there to
get the covariant derivatives of the coincident limits, since
application of Synge's theorem will move derivatives acting at
$y$ such that we also need the covariant derivatives of the
coincident limits. Now that we have the component values of all
the covariant derivatives of the various coincident limits of the
regularized Green function $G_{{\rm ren}}$, we substitute
them in the coincident limit expression
for the noise kernel, Eqn (4.16) of Paper I,
 and arrive at the final result we seek.

Before turning to specific metrics and as a check,
we can reproduce the derivation of the
renormalized stress tensor. We start with the point separated
expression for the stress tensor, which for a massless, conformally
coupled scalar field is \begin{eqnarray}
\left< T_{ab}(x,y) \right> &=&
 \frac{1}{3} \left( g{}^{p'}{}_{b}\,G_{{\rm ren}}{}_{;}{}_{p'}{}_{a} +
  g{}^{p'}{}_{a}\,G_{{\rm ren}}{}_{;}{}_{p'}{}_{b} \right)
-\frac{1}{6} g{}^{p'}{}_{q}\,G_{{\rm ren}}{}_{;}{}_{p'}{}^{q}\,g{}_{a}{}_{b} \cr
&&
-\frac{1}{6} \left( g{}^{p'}{}_{a}\,g{}^{q'}{}_{b}\,G_{{\rm
ren}}{}_{;}{}_{p'}{}_{q'} +
  G_{{\rm ren}}{}_{;}{}_{a}{}_{b} \right)
+\frac{1}{6} \left( \left( G_{{\rm ren}}{}_{;}{}_{p'}{}^{p'} +
    G_{{\rm ren}}{}_{;}{}_{p}{}^{p} \right) \,g{}_{a}{}_{b} \right) \cr
&& +\frac{1}{6} G_{{\rm ren}} \left( {\frac{-\left(
R\,g{}_{a}{}_{b} \right) }{2}} + R{}_{a}{}_{b} \right) \end{eqnarray} We
take the coincident limit and utilize Synge's theorem to obtain
\begin{eqnarray}
\left< T_{ab} \right>_{\rm ren} &=&
 \frac{1}{6} \left( 3\,{\left[G_{{\rm ren}}{}_{;}{}_{a}\right]}{}_{;}{}_{b} +
  3\,{\left[G_{{\rm ren}}{}_{;}{}_{b}\right]}{}_{;}{}_{a} -
  {\left[G_{{\rm ren}}\right]}{}_{;}{}_{a}{}_{b} -
  6\,{\left[G_{{\rm ren}}{}_{;}{}_{a}{}_{b}\right]} \right)
\cr &&
-\frac{1}{6} \left( 3\,{\left[G_{{\rm ren}}{}_{;}{}_{p}\right]}{}^{;}{}^{p} -
    {\left[G_{{\rm ren}}\right]}{}_{;}{}_{p}{}^{p} -
    3\,{\left[G_{{\rm ren}}{}_{;}{}_{p}{}^{p}\right]} \right) \,g{}_{a}{}_{b}
\cr &&
-\frac{1}{12} {\left[G_{{\rm ren}}\right]}\,\left( R\,g{}_{a}{}_{b} -
2\,R{}_{a}{}_{b}
     \right)  \nonumber \\
\cr
&=&
-\frac{1}{6} \left( G_{{\rm ren}}^{(0)}{}_{;}{}_{a}{}_{b} + 12\,G_{{\rm
ren}}^{(2)}{}_{a}{}_{b} -
  G_{{\rm ren}}^{(0)}\,R{}_{a}{}_{b} \right)
\cr &&
-\frac{1}{12}  \left( G_{{\rm ren}}^{(0)}\,R - 2\,G_{{\rm
ren}}^{(0)}{}_{;}{}_{p}{}^{p} -
    12\,G_{{\rm ren}}^{(2)}{}_{p}{}^{p} \right) \,g{}_{a}{}_{b}
\end{eqnarray}
With the explicit values (\ref{Opt-Gvalues2}) for the expansion tensors
we recover the familiar result:
\begin{equation}
\left< T_{ab} \right>_{\rm ren} =
{\frac{{{\kappa}^4}}{1440\,{{\pi }^2}}} \left(
  g{}_{a}{}_{b}
  -4\,{\delta}{}_{a}{}^{\tau}\,{\delta}{}_{b}{}^{\tau}
\right) \end{equation}

%%%%%%%%%%%%%%%%%%%%%%%%%%%%%%%%%%%%%%%%%%%%%%%

\section{Examples}
\label{sec-Examples}

%%%%%%%%%%%%%%%%%%%%%%%%%%%%%%%%%%%%%%%%%%%%%%%

\subsection{Hot Flat Space} \label{subsec-hotflat}

The first example we consider is that of a finite temperature $T =
\kappa/2\pi$ quantum scalar field in flat space. With this, the
stress tensor takes the usual form \begin{equation} \left< T_{ab} \right> =
{\rm diag} \{ -{\frac{1}{3}},-{\frac{1}{3}},-{\frac{1}{3}},1\}
\rho, \quad \rho = - {\frac{{{\pi }^2}\,{T^4}}{30}} \end{equation} with
$x^a = (x,y,z,\tau)$.

Using (\ref{Opt-flat-co}) and (\ref{Opt-CDGr-to-Grn}), the non-zero
coincident limits of the derivatives of $G_{{\rm ren}}$ are
\begin{subequations}\begin{eqnarray}
{\left[G_{{\rm ren}}\right]} &=& {\frac{{T^2}}{12}} \\
{\left[G_{{\rm ren}}{}_{;}{}_{a}{}_{b}\right]} &=&
  -{\frac{\left( {{\pi }^2}\,{T^4} \right) }{90}} \left(
        {\delta}{}_{a}{}_{b} -
4\,{\delta}{}_{a}{}^{\tau}\,{\delta}{}_{b}{}^{\tau}
\right)
\\
{\left[G_{{\rm ren}}{}_{;}{}_{a}{}_{b}{}_{c}{}_{d}\right]} &=&
{\frac{4\,{{\pi }^4}\,{T^6}}{315}} \left(
    {\delta}{}_{a}{}_{b}\,{\delta}{}_{c}{}_{d} -
  12\,{\delta}{}_{a}{}^{\tau}\,{\delta}{}_{b}{}^{\tau}\,
   {\delta}{}_{c}{}_{d} + 16\,{\delta}{}_{a}{}^{\tau}\,
   {\delta}{}_{b}{}^{\tau}\,{\delta}{}_{c}{}^{\tau}\,{\delta}{}_{d}{}^{\tau}
\right)
\end{eqnarray}\end{subequations}
Taking the coincident limit of the massless case for
the noise kernel, Eqn (4.16) of Paper I,
(for this case, we keep
$\xi$ arbitrary), we find
\begin{eqnarray}
N_{abcd} &=& {\frac{{{\pi }^4}\,{T^8}}{226800}} \left\{
 32\,\left( 7 - 28\,\xi + 162\,{{\xi}^2} \right) \,g{}^{\tau}{}_{a}\,
  g{}^{\tau}{}_{b}\,g{}^{\tau}{}_{c}\,g{}^{\tau}{}_{d} \right. \cr
&& -8\,\left( 7 - 42\,\xi + 130\,{{\xi}^2} \right) \,
  \left( g{}_{c}{}_{d}\,g{}^{\tau}{}_{a}\,g{}^{\tau}{}_{b} +
    g{}_{a}{}_{b}\,g{}^{\tau}{}_{c}\,g{}^{\tau}{}_{d} \right)  \cr
&& \hspace{-5mm}
  -4\,\left( 7 - 28\,\xi + 148\,{{\xi}^2} \right) \,
  \left( g{}_{b}{}_{d}\,g{}^{\tau}{}_{a}\,g{}^{\tau}{}_{c} +
    g{}_{a}{}_{d}\,g{}^{\tau}{}_{b}\,g{}^{\tau}{}_{c} +
    g{}_{b}{}_{c}\,g{}^{\tau}{}_{a}\,g{}^{\tau}{}_{d} +
    g{}_{a}{}_{c}\,g{}^{\tau}{}_{b}\,g{}^{\tau}{}_{d} \right)
\cr &&  \hspace{-5mm} \left.
  +4\,\left( 7 - 63\,\xi + 167\,{{\xi}^2} \right)
\,g{}_{a}{}_{b}\,g{}_{c}{}_{d} + \left( 7 - 28\,\xi + 108\,{{\xi}^2} \right) \,
  \left( g{}_{a}{}_{d}\,g{}_{b}{}_{c} + g{}_{a}{}_{c}\,g{}_{b}{}_{d} \right)
\right\} \end{eqnarray}
%@(This expression should be compared with the
%noise kernel expression obtained by Campos and Hu \cite{CamHu}
%for a thermal field in a weak gravitational field background.
%Theirs is more complex in that their calculation includes
%self-consistent backreaction between the scalar and the
%gravitational field while here it is a simple test field
%calculation.)@ -- this is done in the next paper for conformally-optical metrics.
From this, we can immediately compute the trace:
\begin{equation} N_p{}^p{}_q{}^q =
 {\frac{{{\pi }^4}\,{T^8}}{1350}} {{\left( -1 + 6\,\xi \right) }^2}
\end{equation} which we see vanishes for the conformal coupling case
($\xi=1/6$). The non-vanishing components for general coupling are
\begin{equation}
\begin{array}{rccccl}
N_{\tau\tau\tau\tau} = &
  {\frac{{{\pi }^4}\,{T^8}}{37800}}
   \left( 7 - 14\,\xi + 270\,{{\xi}^2} \right)
&   \stackrel{\xi\goesto 0}{\longrightarrow}  &
   {\frac{{{\pi }^4}\,{T^8}}{5400}}
& = &
 {\frac{1}{6}}
  \left< T_{\tau\tau} \right>^2 \\
N_{xxxx} =&
  {\frac{{{\pi }^4}\,{T^8}}{113400}}
   \left( 21 - 154\,\xi + 442\,{{\xi}^2} \right)
&   \stackrel{\xi\goesto 0}{\longrightarrow}  &
   {\frac{{{\pi }^4}\,{T^8}}{5400}}
& = &
 {\frac{3}{2}}
  \left< T_{xx} \right>^2 \\
N_{\tau\tau xx} = & -
  {\frac{{{\pi }^4}\,{T^8}}{56700}}
   \left( 7 - 21\,\xi + 93\,{{\xi}^2} \right)
&   \stackrel{\xi\goesto 0}{\longrightarrow}  &
   -{\frac{{{\pi }^4}\,{T^8}}{8100}}
& = &
 {\frac{1}{3}}
  \left< T_{\tau\tau} \right> \left< T_{xx} \right> \\
N_{xxyy} =&
  {\frac{{{\pi }^4}\,{T^8}}{56700}}
   \left( 7 - 63\,\xi + 167\,{{\xi}^2} \right)
&   \stackrel{\xi\goesto 0}{\longrightarrow}  &
   {\frac{{{\pi }^4}\,{T^8}}{8100}}
& = &
 1
  \left< T_{xx} \right>^2  \\
N_{\tau x\tau x} =& -
  {\frac{{{\pi }^4}\,{T^8}}{226800}}
   \left( 21 - 84\,\xi + 484\,{{\xi}^2} \right)
&   \stackrel{\xi\goesto 0}{\longrightarrow} &
   -{\frac{{{\pi }^4}\,{T^8}}{10800}}
& = &
 {\frac{1}{4}}
  \left< T_{\tau\tau} \right>  \left< T_{xx} \right> \\
N_{xyxy} =&
  {\frac{{{\pi }^4}\,{T^8}}{226800}}
   \left( 7 - 28\,\xi + 108\,{{\xi}^2} \right)
&   \stackrel{\xi\goesto 0}{\longrightarrow}  &
   {\frac{{{\pi }^4}\,{T^8}}{32400}}
& = &
 {\frac{1}{4}}
  \left< T_{xx} \right>^2
\end{array}
\end{equation}
and those that follow from the symmetry of the metric.

We use the dimensionless measure of fluctuations
\cite{KuoFor,PH97,PH0,WuFor}(this is not a tensor, but a measure
of the fluctuations for each component): \begin{equation} \Delta_{abcd} =
\left|
  \frac{ \left< T_{ab} T_{cd} \right> -
     \left< T_{ab}\right> \left< T_{cd} \right>}
       { \left< T_{ab} T_{cd} \right> }
                \right|
 = \left| \frac{ 4 N_{abcd} }
    {4N_{abcd} + \left< T_{ab}\right> \left< T_{cd} \right> }
   \right|
\label{def-Delta-measure} \end{equation} From inspection, $0 \le
\Delta_{abcd} \le 1$. Only for $\Delta \sim 0$ can the
fluctuations be viewed as small. For $\Delta \sim 1$  the
fluctuations are comparable to the mean value.

For hot flat space, the  results for $\Delta$ are \begin{equation}
\begin{array}{cccc}
abcd & \Delta_{abcd} & \xi = 0 & \xi = \frac{1}{6} \\
\tau\tau\tau\tau &
   {\frac{2\,\left( 7 - 14\,\xi + 270\,{{\xi}^2} \right) }
   {35 - 28\,\xi + 540\,{{\xi}^2}}} &
   {\frac{2}{5}} \sim 0.4 &
   {\frac{73}{136}} \sim 0.54 \\
  xxxx & {\frac{2\,\left( 21 - 154\,\xi + 442\,{{\xi}^2} \right) }
   {49 - 308\,\xi + 884\,{{\xi}^2}}} &
   {\frac{6}{7}} \sim 0.86 &
   {\frac{137}{200}} \sim 0.69 \\
  \tau\tau xx & {\frac{4\,\left( 7 - 21\,\xi + 93\,{{\xi}^2} \right) }
   {49 - 84\,\xi + 372\,{{\xi}^2}}} &
   {\frac{4}{7}} \sim 0.57 &
   {\frac{73}{136}} \sim 0.54 \\
  xxyy & {\frac{4\,\left( 7 - 63\,\xi + 167\,{{\xi}^2} \right) }
   {35 - 252\,\xi + 668\,{{\xi}^2}}} &
   {\frac{4}{5}} \sim 0.8 &
   {\frac{41}{104}} \sim 0.39
\end{array}
\label{Opt-hotflat-results} \end{equation} From these results we see, even
for the simple case of thermal fluctuations in flat space, the
fluctuations present in the stress tensor are important.
Discussions on the implication of the fluctuation to mean ratio
can be found in \cite{KuoFor,HP0,PH0,WuFor,ForWu}.

%%%%%%%%%%%%%%%%%%%%%%%%%%%%%%%%%%%%%%%%%%%%%%

\subsection{Optical Schwarzschild black hole}
\label{subsec-opt-schw}

%%%%%%%%%%%%%%%%%%%%%%%%%%%%%%%%%%%%%%%%%%%

We now consider the optical spacetime conformally related to the
Schwarzschild black hole spacetime. For this spacetime, the line
element is \begin{equation} ds^2 = d\tau^2 + \left(1-\frac{2M}{r}\right)^{-2}
dr^2 +
  \left(1-\frac{2M}{r}\right)^{-1} r^2
  \left(d\theta^2+\sin^2\theta d\phi^2\right)
\label{Opt-schw-metricg} \end{equation} Taking $\kappa=2\pi T$ and
$T=1/(8\pi M)$ we choose the quantum state
corresponding to the Hartle-Hawking state in the
conformally-related Schwarzschild spacetime.
We  use the spacetime coordinates $x^a = (r,\theta,\phi,\tau)$
and introduce the rescaled inverse radial coordinate
$x=2M/r = 1/(4\pi T r)$. Spatial infinity corresponds to $x=0$
 and $x=1$ is the black hole horizon. For a
massless conformal scalar field ($m=0$, $\xi=1/6$) the stress tensor
is \begin{equation} \left< T_a{}^b \right> = {\rm diag} \left\{
  {\frac{1}{3}}, {\frac{1}{3}}, {\frac{1}{3}}, -1 \right\}
   {\frac{{{\pi }^2}\,{T^4}}{30}}
\end{equation}
We recover the standard thermal result for the stress tensor.
The component values for the noise kernel are
\begin{subequations}\begin{eqnarray}
N_\tau{}^\tau{}_\tau{}^\tau &=&   {\frac{{{\pi }^4}\,{T^8}}{2041200}} \left(
    657 - 1050\,{x^4} + 8400\,{x^6} - 16800\,{x^7} - 16997400\,{x^8}
\right. \cr  && \hspace{7mm}\left.
 + 80206000\,{x^9}
 -140910000\,{x^{10}} + 109242000\,{x^{11}} -  31515750\,{x^{12}}
\right)
\\
N_r{}^r{}_r{}^r &=&   {\frac{{{\pi }^4}\,{T^8}}{2041200}} \left( 137 +
1120\,{x^3} - 210\,{x^4} + 80640\,{x^5} - 103600\,{x^6} + 16800\,{x^7} \right.
\cr
  && \hspace{7mm} \left.+ 9829800\,{x^8} - 43285200\,{x^9} +
73602000\,{x^{10}} - 57186000\,{x^{11}} \right. \cr
  && \hspace{7mm} \left.+ 17057250\,{x^{12}} \right)
\\
N_\theta{}^\theta{}_\theta{}^\theta &=&  {\frac{{{\pi }^4}\,{T^8}}{2041200}}
\left( 137 - 560\,{x^3} + 1470\,{x^4} + 30240\,{x^5} - 44800\,{x^6} -
16800\,{x^7} \right. \cr
  && \hspace{7mm} \left. -15805800\,{x^8} + 73203600\,{x^9} -
127386000\,{x^{10}} + 98820000\,{x^{11}} \right. \cr
  && \hspace{7mm} \left. -28788750\,{x^{12}} \right)
\\
N_\tau{}^\tau{}_r{}^r &=&   {\frac{{{\pi }^4}\,{T^8}}{6123600}} \left( -657 +
6720\,{x^3} - 5670\,{x^4} - 42000\,{x^6} + 484920\,{x^8} +
  4202480\,{x^9} \right. \cr
  && \hspace{7mm} \left. -12514800\,{x^{10}} + 9439200\,{x^{11}} -
1548450\,{x^{12}} \right)
\\
N_\tau{}^\tau{}_\theta{}^\theta &=&   {\frac{{{\pi }^4}\,{T^8}}{6123600}}
\left( -657 - 3360\,{x^3} + 4410\,{x^4} + 8400\,{x^6} + 25200\,{x^7} -
  53478360\,{x^8} \right. \cr
  && \hspace{4mm} \left.+ 248828960\,{x^9} - 434589600\,{x^{10}} +
337051800\,{x^{11}} -
  97825050\,{x^{12}} \right)
\\
N_r{}^r{}_\theta{}^\theta &=&  {\frac{{{\pi }^4}\,{T^8}}{6123600}} \left( 123
- 5040\,{x^3} + 3150\,{x^4} - 120960\,{x^5} + 176400\,{x^6} - 25200\,{x^7}
\right. \cr
  && \hspace{7mm} \left.+ 2858760\,{x^8} - 10502560\,{x^9} +
15885600\,{x^{10}} - 12447000\,{x^{11}} \right. \cr
  && \hspace{7mm} \left.+ 4178250\,{x^{12}} \right)
\label{Opt-schw-complete-comp}
\end{eqnarray}\end{subequations}

From the component values of the noise kernel we can compute its
trace \begin{equation} N = N_p{}^p{}_q{}^q = - {\frac{4\,{{\pi
}^4}\,{T^8}\,{x^8}}{567}}
        \left( 9720 - 45832\,x + 80520\,{x^2} - 62424\,{x^3} + 18009\,{x^4}
\right) \label{Opt-noise-trace} \end{equation} We know from prior results
%Section (\ref{sec-ps-trace})
that this should vanish, since we have worked with a massless
conformally coupled scalar field. This failure of the trace of
the noise kernel to vanish is due to the failure of the Gaussian
approximated Green function (\ref{Chap4-GGauss}) to satisfy the
field equation to fourth order.

\section{Failure of Gaussian Approximation at the Fourth Order}
\label{sec-errGauss}

To be sure that this error does not arise from the symbolic
manipulation, let us mention ways to check the correctness of the
algorithm. The basic procedures for generating the needed series
expansions are recursive on the expansion order. (The recursion
formulas of the expansion used in point-separation are collected
in the Appendices). For the noise kernel we need results up to
fourth order in the separation distance. The well established
work for the stress tensor is to second order. This provides a
check of our code by verifying we always get the known results
for the stress tensor expectation value. Once we know the second
order recursion is correct, we know the algorithm is functioning
as desired. The correctness in the (new) fourth order terms
becomes particularly important in the work in Paper III, when we
consider metrics conformally related, as we get intermediate
results of up to 1100 terms in length.

We can check the accuracy of the Gaussian approximation by using
the computed component values of coincident limit of the
covariant derivatives of $G_{\rm ren}$. We have assumed \begin{equation}
G_{\rm ren}{}_{;}{}_{p}{}^{p} - {\frac{G_{\rm ren}\,R}{6}} = 0
\label{Opt-feqn} \end{equation} This can be used to test the approximation
order by order. To test up to the second order, we just take the
coincident limit
\begin{equation} {\left[G_{\rm
ren}{}_{;}{}_{p}{}^{p}\right]} - {\frac{R\,{\left[G_{\rm
ren}\right]}}{6}} = 0. \label{Opt-feqn-o2}
\end{equation}
For the metric
(\ref{Opt-schw-metricg}), the scalar curvature is \begin{equation} R = -
{\frac{6\,{M^2}}{{r^4}}} = - 24\,{{\pi }^2}\,{T^2}\,{x^4} \end{equation}
while the results of our computation of the component values of
the coincident limit of the covariant derivatives of $G_{\rm ren}$
yield \begin{equation} {\left[G_{\rm ren}\right]} = {\frac{4\,{{\pi
}^2}\,{T^2}}{3}}, \quad {\rm and} \quad {\left[G_{\rm
ren}{}_{;}{}_{p}{}^{p}\right]} = - {\frac{16\,{{\pi
}^4}\,{T^4}\,{x^4}}{3}}. \end{equation} With these values, Eq.
(\ref{Opt-feqn-o2}) can be seen to be satisfied. Thus the
Gaussian approximation is good to the second order. (This had
better be the case, as the approximation at this order has been
checked against numerical computations of the stress tensor by
other authors. See e.g., description in \cite{AHL,HLA})

To check the third order term, we take one covariant derivative and then the
coincident limit of (\ref{Opt-feqn}):
\begin{equation}
{\left[G_{\rm ren}{}_{;}{}_{p}{}^{p}{}_{a}\right]}
-\frac{1}{6}\left( R{}_{;}{}_{a}\,{\left[G_{\rm ren}\right]} +
  R\,{\left[G_{\rm ren}{}_{;}{}_{a}\right]} \right) = 0
\label{Opt-feqn-o3}
\end{equation}
Using the results ${\left[G_{\rm ren}{}_{;}{}_{a}\right]} = 0$,
\begin{equation}
R_{;a}  = \left\{ 384\,{{\pi }^3}\,{T^3}\,{x^5},0,0,0\right\}
\end{equation}
and
\begin{equation}
{\left[G_{\rm ren}{}_{;}{}_{p}{}^{p}{}_{a}\right]} = \left\{ {\frac{256\,{{\pi
}^5}\,{T^5}\,{x^5}}{3}}
                     ,0,0,0\right\}
\end{equation} we see Eq. (\ref{Opt-feqn-o3}) is also satisfied. This has to
be the case: (\ref{Opt-Gvalues2}) shows the third order expansion
tensor does not have any contribution from the $W(x,y)$ part of
the Hadamard ansatz and this is the only place a lack of symmetry
in $G_{\rm ren}(x,y)$ could come in. Therefore the Gaussian Green
function (\ref{Chap4-GGauss}) is symmetric. For symmetric
functions, the odd order expansion tensors are determined
completely by the even order tensors (see
(\ref{apx-endpoint-odd3})).

Continuing to fourth order, (\ref{Opt-feqn}) becomes
\begin{eqnarray}
{\left[G_{\rm ren}{}_{;}{}_{p}{}^{p}{}_{a}{}^{b}\right]} -\frac{1}{6}\left(
R{}_{;}{}_{a}{}^{b}\,{\left[G_{\rm ren}\right]} +
  R{}^{;}{}^{b}\,{\left[G_{\rm ren}{}_{;}{}_{a}\right]} +
  R{}_{;}{}_{a}\,{\left[G_{\rm ren}{}^{;}{}^{b}\right]} +
  R\,{\left[G_{\rm ren}{}_{;}{}_{a}{}^{b}\right]} \right) &=& \cr
{\left[G_{\rm ren}{}_{;}{}_{p}{}^{p}{}_{a}{}^{b}\right]} -\frac{1}{6}\left(
R{}_{;}{}_{a}{}^{b}\,{\left[G_{\rm ren}\right]} +
  R\,{\left[G_{\rm ren}{}_{;}{}_{a}{}^{b}\right]} \right) &=& 0\cr
&&
\label{Opt-feqn-o4}
\end{eqnarray}
Proceeding as before and evaluating the left hand side above,
the component values are
\begin{eqnarray}
{\rm diag}\left\{\right.&
  -{\frac{128\,{{\pi }^6}\,{T^6}\,{x^8}}{315}} \left( 162648 - 746888\,x +
1295880\,{x^2} - 1005480\,{x^3} + 293805\,{x^4} \right), \cr &
   {\frac{128\,{{\pi }^6}\,{T^6}\,{x^8}}{315}} \left( 8424 - 29704\,x +
44040\,{x^2} - 34560\,{x^3} + 11835\,{x^4} \right), \cr &
   {\frac{128\,{{\pi }^6}\,{T^6}\,{x^8}}{315}} \left( 8424 - 29704\,x +
44040\,{x^2} - 34560\,{x^3} + 11835\,{x^4} \right), \cr &
   {\frac{128\,{{\pi }^6}\,{T^6}\,{x^8}}{189}} \left( 9720 - 45832\,x +
80520\,{x^2} - 62424\,{x^3} + 18009\,{x^4} \right)
\left. \right\}
\end{eqnarray}
The failure of the left hand side of (\ref{Opt-feqn-o4}) to vanish
shows that the failure of the trace to vanish comes from the
limitations of the
Gaussian approximation. The Gaussian approximation is only useful up to the
third order in $\sigma^a$.

With this knowledge, the trace (\ref{Opt-noise-trace}) becomes
our measure of the error in the noise kernel from the use of the
Gaussian approximation. It is important to note that the noise
kernel trace $N$ vanishes as $x\goesto 0$, or, $r\goesto\infty$,
i.e. where one would expect the effects of curvature to vanish.
We can also see from (\ref{Opt-schw-complete-comp}) that the
noise is finite at the horizon ($x=1$).

Using our derived expression for the noise kernel we see that its
trace vanishes at spatial infinity, thus we can trust our results
there. Using the measure (\ref{def-Delta-measure}), the
fluctuations at $r\goesto\infty$ are \begin{equation}
\begin{array}{ccccccc}
abcd: &
\tau\tau\tau\tau & rrrr & \theta\theta\theta\theta &
\tau\tau rr & \tau\tau\theta\theta & rr\theta\theta \\
\Delta_{abcd}: &
  {\frac{73}{136}} &
  {\frac{137}{200}} &
  {\frac{137}{200}} &
  {\frac{73}{136}} &
  {\frac{73}{136}} &
  {\frac{41}{104}}
\end{array}
\end{equation} which match exactly the results (\ref{Opt-hotflat-results})
for hot flat space with conformal coupling, another reassuring
fact. Since the computation of the noise kernel for metric
(\ref{Opt-schw-metricg}) is much more involved, this provides yet
another check of our symbolic computer code.

Now having shown that it is truly the Gaussian approximation that
is at fault for the failure of the noise kernel trace to vanish,
we can use its value as a measure of the error in the results
(\ref{Opt-schw-complete-comp}). Since $N$ should be zero,
$N/N_{abcd}$ is a dimensionless measure of this error. At the
horizon ($x=1$), \begin{equation}
\begin{array}{ccccccc}
abcd: &
\tau\tau\tau\tau & rrrr & \theta\theta\theta\theta &
\tau\tau rr & \tau\tau\theta\theta & rr\theta\theta \\
\frac{N_p{}^p{}_q{}^q}{N_a{}^b{}_c{}^d}: &
  627 \% &
  791 \% &
  791 \% &
  1390 \% &
  1390 \% &
  19855 \%
\end{array}
\end{equation} These errors show the Gaussian approximation fails to
provide reliable results for the noise kernel near the horizon of
the optical Schwarzschild metric. We expect this be the case also
for the Schwarzschild metric near the horizon, an explicit
calculation will appear in the next paper. In the above we have
identified the occurrence of significant error begins at the
fourth covariant derivative order.

In conclusion, towards the three goals set for this paper, we
have detailed the steps in implementing the modified point
separation scheme  under the Gaussian approximation for the Green
function for a quantum scalar field in the optical spacetimes. We
have derived the regularized noise kernel for a thermal field in
flat space. We have obtained a finite expression for the noise
kernel at the horizon of an optical Schwarzschild spacetime and
recovered the hot flat space result at infinity. From the error
in the noise kernel at the horizon we showed that the Gaussian
approximation scheme of Bekenstein-Parker-Page applied to the
Green function which provides surprisingly good results for the
stress tensor involving the second covariant derivative order of
the Green function, fails at the fourth covariant derivative order. 

\begin{acknowledgments}
NGP thanks Professor R. M. Wald for
correspondence on his  regularization procedures, and Professors
S. Christensen and L. Parker for the use of the MathTensor
program. BLH thanks Professors Paul Anderson, Sukanya Sinha and E.
Verdaguer for discussions. This work is supported in part by NSF
grant PHY98-00967.
\end{acknowledgments}

%%%%%%%%%%%%%%%%%%%%%%%%%%%%%%%%%%%%%%%%%%%%%%%%
\newpage
\appendix

%%%%%%%%%%%%%%%%%%%%%%%%%%%%%%%%%%%%%%%%%%%%%%%

\section{World function $\sigma$}
\label{apx-sigma}

In this appendix, we review the properties of the world function
$\sigma$. We also demonstrate how symbolic computations are
implemented and used in this work.
Christensen's \cite{ChristWinnipeg} method for
determining the coincident limit of covariant derivatives of
functions defined via a covariant differential equation is
reviewed.

The world function is defined to be one half of the square of the
geodesic distance between two points on a differential manifold.
As such, it satisfies the equation \begin{equation} \sigma -
{\frac{\sigma{}_{;}{}_{p}\,\sigma{}^{;}{}^{p}}{2}} = 0
\label{PSApx-sigmadef} \end{equation} along with the initial value \begin{equation}
{\left[\sigma{}_{;}{}_{a}\right]} = 0 \label{PSApx-sigmaone} \end{equation}

To determine $\left[ \sigma_{;ab} \right]$, we take two
derivatives of (\ref{PSApx-sigmadef}). To help illustrate how
these calculations are done on the computer, the output presented
here is direct output from MathTensor\cite{MathTensor},
with Mathematica\cite{Mathematica} carrying
out the formatting. We use the MathTensor function {\tt CD} and
get \begin{equation} \sigma{}_{;}{}_{a}{}_{b} -
{\frac{\sigma{}_{;}{}_{p}{}_{b}\,
      \sigma{}_{;}{}_{a}{}^{p}}{2}} -
  {\frac{\sigma{}_{;}{}_{p}{}_{a}\,\sigma{}_{;}{}_{b}{}^{p}}{2}} -
  {\frac{\sigma{}^{;}{}^{p}\,\sigma{}_{;}{}_{p}{}_{a}{}_{b}}{2}} -
  {\frac{\sigma{}_{;}{}_{p}\,\sigma{}_{;}{}_{a}{}^{p}{}_{b}}{2}} = 0
\end{equation} This is put into a canonical form via {\tt Canonicalize} \begin{equation}
\sigma{}_{;}{}_{a}{}_{b} - \sigma{}_{;}{}_{p}{}_{a}\,
   \sigma{}_{;}{}_{b}{}^{p} - \sigma{}_{;}{}_{p}\,
   \sigma{}_{;}{}_{a}{}^{p}{}_{b} = 0
\end{equation} The condition (\ref{PSApx-sigmaone}) is encoded into
Mathematica by defining
a rule \\
{\tt Ci[ CD[sigma,la\_] ] -> 0}, where ${\tt Ci[]}$ is a function
defined to represent the coincident limit and is formatted to be
displayed using the standard $[\cdots]$ notation. Using this
rule, the coincident limit is \begin{equation}
{\left[\sigma{}_{;}{}_{a}{}_{b}\right]} -
  {\left[\sigma{}_{;}{}_{p}{}_{a}\right]}\,
   {\left[\sigma{}_{;}{}_{b}{}^{p}\right]} = 0
\end{equation} This immediately shows
${\left[\sigma{}_{;}{}_{a}{}_{b}\right]} = g{}_{a}{}_{b}$ and
such a rule is defined.

Proceeding with one more covariant derivative, \begin{equation} 0 =
\sigma{}_{;}{}_{a}{}_{b}{}_{c} -
  \sigma{}_{;}{}_{p}{}_{c}\,\sigma{}_{;}{}_{a}{}^{p}{}_{b} -
  \sigma{}_{;}{}_{p}{}_{b}\,\sigma{}_{;}{}_{a}{}^{p}{}_{c} -
  \sigma{}_{;}{}_{p}{}_{a}\,\sigma{}_{;}{}_{b}{}^{p}{}_{c} -
  \sigma{}_{;}{}_{p}\,\sigma{}_{;}{}_{a}{}^{p}{}_{b}{}_{c},
\end{equation} and using the two rules we already have,
we recursively get
\begin{equation} 0 =
{\left[\sigma{}_{;}{}_{a}{}_{b}{}_{c}\right]} +
  {\left[\sigma{}_{;}{}_{a}{}_{c}{}_{b}\right]}
\end{equation}
Using MathTensors'
{\tt OrderCD}, which commutes the covariant derivatives on each
term until they are in alphabetical order, we get the result
\begin{eqnarray} 0 &=&
2\,{\left[\sigma{}_{;}{}_{a}{}_{b}{}_{c}\right]} +
  {\left[\sigma{}_{;}{}_{p}\right]}\,R{}_{a}{}^{p}{}_{b}{}_{c}\cr
  &=& 2\,{\left[\sigma{}_{;}{}_{a}{}_{b}{}_{c}\right]}
\end{eqnarray} where (\ref{PSApx-sigmaone}) is used to go from the first
to the second line. We have the coincident limit of three
covariant derivatives acting on $\sigma$ vanishes.

Proceeding with one more covariant derivative: 
\begin{eqnarray} 0 &=&
\sigma{}_{;}{}_{p}{}_{a}{}_{d}\,\sigma{}_{;}{}_{b}{}^{p}{}_{c} +
  \sigma{}_{;}{}_{p}{}_{a}{}_{c}\,\sigma{}_{;}{}_{b}{}^{p}{}_{d} +
  \sigma{}_{;}{}_{p}{}_{a}{}_{b}\,\sigma{}_{;}{}_{c}{}^{p}{}_{d} -
  \sigma{}_{;}{}_{a}{}_{b}{}_{c}{}_{d} +
  \sigma{}_{;}{}_{p}{}_{d}\,\sigma{}_{;}{}_{a}{}^{p}{}_{b}{}_{c} \cr
&&        +
\sigma{}_{;}{}_{p}{}_{c}\,\sigma{}_{;}{}_{a}{}^{p}{}_{b}{}_{d} +
  \sigma{}_{;}{}_{p}{}_{b}\,\sigma{}_{;}{}_{a}{}^{p}{}_{c}{}_{d} +
  \sigma{}_{;}{}_{p}{}_{a}\,\sigma{}_{;}{}_{b}{}^{p}{}_{c}{}_{d} +
  \sigma{}_{;}{}_{p}\,\sigma{}_{;}{}_{a}{}^{p}{}_{b}{}_{c}{}_{d} \cr
&\Rightarrow& \cr 0 &=&
{\left[\sigma{}_{;}{}_{a}{}_{b}{}_{c}{}_{d}\right]} +
  {\left[\sigma{}_{;}{}_{a}{}_{c}{}_{b}{}_{d}\right]} +
  {\left[\sigma{}_{;}{}_{a}{}_{d}{}_{b}{}_{c}\right]}
\end{eqnarray} where once again we use the rules we already know. Now
commuting the covariant derivatives \begin{eqnarray} 0 &=&
3\,{\left[\sigma{}_{;}{}_{a}{}_{b}{}_{c}{}_{d}\right]} + \left(
R{}_{a}{}^{p}{}_{b}{}_{c}{}_{;}{}_{d} +
    R{}_{a}{}^{p}{}_{b}{}_{d}{}_{;}{}_{c} \right) \,
  {\left[\sigma{}_{;}{}_{p}\right]} \cr
  && + {\left[\sigma{}_{;}{}_{p}{}_{d}\right]}\,R{}_{a}{}^{p}{}_{b}{}_{c} +
  {\left[\sigma{}_{;}{}_{p}{}_{c}\right]}\,R{}_{a}{}^{p}{}_{b}{}_{d} +
  {\left[\sigma{}_{;}{}_{p}{}_{b}\right]}\,R{}_{a}{}^{p}{}_{c}{}_{d} +
  {\left[\sigma{}_{;}{}_{p}{}_{a}\right]}\,R{}_{b}{}^{p}{}_{c}{}_{d}
\end{eqnarray} and using the known rules, we get the equation
\begin{equation} 0 =
3\,{\left[\sigma{}_{;}{}_{a}{}_{b}{}_{c}{}_{d}\right]} +
  R{}_{a}{}_{c}{}_{b}{}_{d} + R{}_{a}{}_{d}{}_{b}{}_{c}
\end{equation}
which we can solve for the term we
need. In practice,
Mathematica's function {\tt Solve[]}
is used, giving us the rule
\begin{equation}
{\left[\sigma{}_{;}{}_{a}{}_{b}{}_{c}{}_{d}\right]} \rightarrow
-\frac{1}{3} \left(R{}_{a}{}_{c}{}_{b}{}_{d} +
R{}_{a}{}_{d}{}_{b}{}_{c} \right) \end{equation}

By knowing the coincident limit of $n-1$ covariant derivatives of
$\sigma$, we determine the coincident limit of $n$ covariant
derivatives. This is the recursive algorithm developed by
Christensen. It is the main idea we use for computing the
expansions needed in this work. A general outline starts by
assuming we have the rules for $n-1$ covariant derivatives, then,
\begin{enumerate}
  \item Take $n$ covariant derivatives of the defining equation
         (in this case, Eq (\ref{PSApx-sigmadef}));
  \item Use the rules for $n-1$ covariant derivatives to get the
        coincident limit;
  \item Commute the covariant derivatives;
  \item Use again the rules for $n-1$ covariant derivatives on the
        terms generated;
  \item Solve for the coincident limit of the $n$ derivative term;
  \item Define a new rule for this term.
\end{enumerate}
These steps are iterated until all terms needed are generated.

For the world function $\sigma$, we need to carry this out to
eight covariant derivatives. The seventh and eighth order
(derivative) terms become quite large. In fact, when computing
these expression, we only substitute (steps 2 and 4 above) up to
four covariant derivatives and still get results with 240 terms
for the seven derivative result and 1101 for the eight. We only
finish carrying out the recursion when we use these highest order
values.

The results for five and six covariant derivatives are:
%\begin{subequations}
\begin{equation} {\left[\sigma{}_{;}{}_{a}{}_{b}{}_{c}{}_{d}{}_{e}\right]} =
-\frac{1}{4} \left( R{}_{a}{}_{c}{}_{b}{}_{d}{}_{;}{}_{e} +
  R{}_{a}{}_{c}{}_{b}{}_{e}{}_{;}{}_{d} +
  R{}_{a}{}_{d}{}_{b}{}_{c}{}_{;}{}_{e} +
  R{}_{a}{}_{d}{}_{b}{}_{e}{}_{;}{}_{c} +
  R{}_{a}{}_{e}{}_{b}{}_{c}{}_{;}{}_{d} +
  R{}_{a}{}_{e}{}_{b}{}_{d}{}_{;}{}_{c} \right)
\end{equation}
\begin{eqnarray}
{\left[\sigma{}_{;}{}_{a}{}_{b}{}_{c}{}_{d}{}_{e}{}_{f}\right]}
&=& -\left(
                   R{}_{a}{}_{c}{}_{b}{}_{d}{}_{;}{}_{e}{}_{f} +
  R{}_{a}{}_{c}{}_{b}{}_{e}{}_{;}{}_{d}{}_{f} +
  R{}_{a}{}_{c}{}_{b}{}_{f}{}_{;}{}_{d}{}_{e} +
  R{}_{a}{}_{d}{}_{b}{}_{c}{}_{;}{}_{e}{}_{f} +
  R{}_{a}{}_{d}{}_{b}{}_{e}{}_{;}{}_{c}{}_{f} \right.    \cr
&&        +        R{}_{a}{}_{d}{}_{b}{}_{f}{}_{;}{}_{c}{}_{e} +
  R{}_{a}{}_{e}{}_{b}{}_{c}{}_{;}{}_{d}{}_{f} +
  R{}_{a}{}_{e}{}_{b}{}_{d}{}_{;}{}_{c}{}_{f} +
  R{}_{a}{}_{e}{}_{b}{}_{f}{}_{;}{}_{c}{}_{d} +
  R{}_{a}{}_{f}{}_{b}{}_{c}{}_{;}{}_{d}{}_{e}       \cr
&&        +\left.
R{}_{a}{}_{f}{}_{b}{}_{d}{}_{;}{}_{c}{}_{e} +
  R{}_{a}{}_{f}{}_{b}{}_{e}{}_{;}{}_{c}{}_{d} \right)/5 \cr
&&        -\left(
R{}_{p}{}_{e}{}_{c}{}_{d}\,R{}_{a}{}_{b}{}_{f}{}^{p} +
  R{}_{p}{}_{e}{}_{b}{}_{d}\,R{}_{a}{}_{c}{}_{f}{}^{p} +
  R{}_{p}{}_{e}{}_{b}{}_{c}\,R{}_{a}{}_{d}{}_{f}{}^{p} +
  R{}_{p}{}_{e}{}_{a}{}_{d}\,R{}_{b}{}_{c}{}_{f}{}^{p} \right. \cr
&&         \left. +
R{}_{p}{}_{e}{}_{a}{}_{c}\,R{}_{b}{}_{d}{}_{f}{}^{p} +
  R{}_{p}{}_{e}{}_{a}{}_{b}\,R{}_{c}{}_{d}{}_{f}{}^{p} \right)/9 \cr
&&        +\left(
R{}_{p}{}_{d}{}_{c}{}_{f}\,R{}_{a}{}_{b}{}_{e}{}^{p} +
  R{}_{p}{}_{d}{}_{c}{}_{e}\,R{}_{a}{}_{b}{}_{f}{}^{p} +
  R{}_{p}{}_{d}{}_{b}{}_{f}\,R{}_{a}{}_{c}{}_{e}{}^{p} +
  R{}_{p}{}_{d}{}_{b}{}_{e}\,R{}_{a}{}_{c}{}_{f}{}^{p} \right. \cr
&&         \left. +
R{}_{p}{}_{d}{}_{a}{}_{f}\,R{}_{b}{}_{c}{}_{e}{}^{p} +
  R{}_{p}{}_{d}{}_{a}{}_{e}\,R{}_{b}{}_{c}{}_{f}{}^{p} \right)/45 \cr
&&        +\left(   R{}_{p}{}_{c}{}_{b}{}_{f}\,\left(
R{}_{a}{}_{d}{}_{e}{}^{p} +
     7\,R{}_{a}{}_{e}{}_{d}{}^{p} \right)  +
  R{}_{p}{}_{c}{}_{b}{}_{e}\,\left( R{}_{a}{}_{d}{}_{f}{}^{p} +
     7\,R{}_{a}{}_{f}{}_{d}{}^{p} \right)       \right. \cr
&&        +         R{}_{p}{}_{c}{}_{b}{}_{d}\,\left(
R{}_{a}{}_{e}{}_{f}{}^{p} +
     7\,R{}_{a}{}_{f}{}_{e}{}^{p} \right)  +
  R{}_{p}{}_{c}{}_{a}{}_{f}\,\left( R{}_{b}{}_{d}{}_{e}{}^{p} +
     7\,R{}_{b}{}_{e}{}_{d}{}^{p} \right)   \cr
&&         \left. + R{}_{p}{}_{c}{}_{a}{}_{e}\,\left(
R{}_{b}{}_{d}{}_{f}{}^{p} +
     7\,R{}_{b}{}_{f}{}_{d}{}^{p} \right)  +
  R{}_{p}{}_{c}{}_{a}{}_{d}\,\left( R{}_{b}{}_{e}{}_{f}{}^{p} +
     7\,R{}_{b}{}_{f}{}_{e}{}^{p} \right)       \right)/45 \cr
&&        +\left(   -\left( R{}_{p}{}_{d}{}_{b}{}_{c}\,
     \left( 5\,R{}_{a}{}_{e}{}_{f}{}^{p} - R{}_{a}{}_{f}{}_{e}{}^{p} \right)
      \right)  - R{}_{p}{}_{d}{}_{a}{}_{c}\,
   \left( 5\,R{}_{b}{}_{e}{}_{f}{}^{p} - R{}_{b}{}_{f}{}_{e}{}^{p} \right)
  \right. \cr
&&        +         2\,R{}_{p}{}_{b}{}_{a}{}_{f}\,\left(
5\,R{}_{c}{}_{d}{}_{e}{}^{p} -
     R{}_{c}{}_{e}{}_{d}{}^{p} \right)  +
  2\,R{}_{p}{}_{b}{}_{a}{}_{e}\,
   \left( 5\,R{}_{c}{}_{d}{}_{f}{}^{p} - R{}_{c}{}_{f}{}_{d}{}^{p} \right)   \cr
&&        +         2\,R{}_{p}{}_{b}{}_{a}{}_{d}\,\left(
5\,R{}_{c}{}_{e}{}_{f}{}^{p} -
     R{}_{c}{}_{f}{}_{e}{}^{p} \right)  -
  R{}_{p}{}_{d}{}_{a}{}_{b}\,\left( 5\,R{}_{c}{}_{e}{}_{f}{}^{p} -
     R{}_{c}{}_{f}{}_{e}{}^{p} \right)   \cr
&&         \left. + 2\,R{}_{p}{}_{b}{}_{a}{}_{c}\,\left(
5\,R{}_{d}{}_{e}{}_{f}{}^{p} -
     R{}_{d}{}_{f}{}_{e}{}^{p} \right)  -
  R{}_{p}{}_{c}{}_{a}{}_{b}\,\left( 5\,R{}_{d}{}_{e}{}_{f}{}^{p} -
     R{}_{d}{}_{f}{}_{e}{}^{p} \right)       \right)/45
\cr && \end{eqnarray}
%\end{subequations}

%%%%%%%%%%%%%%%%%%%%%%%%%%%%%%%%%%%%%%%%%%%%%%%%
\section{End Point Series Expansion}
\label{apx-end-pt-series}

The basic input into the computation of the stress tensor
or noise kernel is the Green function, a perfect example of a
bi-scalar. We want to express it in such a way we can easily identify
how it depends on the distance between its two support points.
This leads us to consider {\em series expansions}
of bi-scalars. The techniques are also useful for the series
expansions of bi-tensors.

The world function $\sigma$ introduced above provides the ideal
geometric object for such a construction. It contains both
distance and direction information. For a bi-scalar $S(x,y)$,
the {\em end point expansion} is \begin{equation} S(x,y) = A^{(0)} + \sigma^p
A^{(1)}_p  + \sigma^p \sigma^q A^{(2)}_{pq} + \cdots +
\sigma^{p_1}\cdots \sigma^{p_n} A^{(n)}_{p_1\cdots p_n} + \cdots,
\label{Apx-series1} \end{equation}
so-called because the expansion tensors
$A^{(n)}_{a_1\cdots a_n} = A^{(n)}_{a_1\cdots a_n}(x)$
have support at one of the end points for which $S(x,y)$ has support.

%An alternate expansion about a point geodesically midway between
%the two points $x$ and $y$ also can be defined, the {\em midpoint
%expansion}. We do not use this expansion for it is not clear how
%to get from the midpoint series expansion to the covariant
%derivatives of the bi-scalar. On the way to the midpoint
%expansion, an expansion is developed about an arbitrary point $z$
%between $x$ and $y$. This obviously greatly increases the
%complexity of the series, as now there are three points with
%which to be concerned, but the derivatives can still be
%retrieved. But when the point $z$ is taken to be geodesic
%midpoint, this distinction is lost and derivatives of the
%midpoint series do not generate the derivatives of the bi-scalar,
%though the series expansion itself greatly simplifies, especially
%for a symmetric bi-scalar. On the other hand, the end point
%series will in general be more involved but easier to manipulate.

It is only the symmetric part
of the expansion tensors $A^{(n)}_{a_1\cdots a_n}$ that
contributes to the expansion, since they are contracted against
symmetric products of $\sigma^{p_i}$'s. Moreover,
the expansion tensors are order $n$ in
distance contribution to the bi-scalar $S(x,y)$. We also find it
convenient to have an expansion where the distance dependence is
separated from the direction dependence. To this end, if $p^a$ is
the unit vector along the geodesic from $x$ to $y$, $\sigma^a =
\left(2\sigma\right)^\frac{1}{2} p^a$, the expansion
(\ref{Apx-series1}) can be re-expressed as \begin{equation} S(x,y) = A^{(0)} +
\sigma^{\frac{1}{2}} A^{(1)} +
        \sigma A^{(2)} + \cdots + \sigma^\frac{n}{2} A^{(n)} + \cdots
\end{equation} where $A^{(n)} = 2^\frac{n}{2} p^{p_1}\cdots p^{p_n}
A^{(n)}_{p_1\cdots p_n}$. Now the expansion scalars $A^{(n)}$
carry the direction information.

When multiplying series, this form readily collects
terms by
their order in distance. In the context of symbolic manipulation
of series on the computer, this alternate form greatly improves
processing speed.

The first series expansion to consider is that for the world
function, which, by virtue of its defining differential equation
is given by \begin{equation} \sigma(x,y) = \frac{g_{pq}}{2}  \sigma^p \sigma^q
\end{equation} This is exact and from it we can see that all expansion
tensors of order $n \ge 3$ vanish. This in turn tells us that
$\left[ \sigma_{;(a_1 a_2 a_3 \cdots a_n)} \right] = 0$, {\it
i.e.}, the coincident limit of three or more symmetrized
covariant derivatives of the world function vanish. This can also
be seen by direct inspection of the previously given expressions
for the coincident limits.

We now turn to relating the expansion tensors to the coincident
limits of the derivatives of the scalar $S$. This is done by
taking covariant derivatives and then the coincident limit of
(\ref{Apx-series1}). We immediately get \begin{equation} {\left[S\right]} =
A^{(0)}. \end{equation} Taking one derivative and the coincident limit gives
\begin{equation} {\left[S{}_{;}{}_{a}\right]} = A^{(1)}{}_{a} +
A^{(0)}{}_{;}{}_{a} \quad \Rightarrow \quad A^{(1)}{}_{a} =
-{\left[S\right]}{}_{;}{}_{a} + {\left[S{}_{;}{}_{a}\right]}. \end{equation}
Two covariant derivatives yields \begin{equation}
{\left[S{}_{;}{}_{a}{}_{b}\right]} =
        A^{(2)}{}_{a}{}_{b} + A^{(2)}{}_{b}{}_{a} + A^{(1)}{}_{a}{}_{;}{}_{b} +
  A^{(1)}{}_{b}{}_{;}{}_{a} + A^{(0)}{}_{;}{}_{a}{}_{b}.
\end{equation} We only need the symmetric part of $A^{(2)}{}_{a}{}_{b}$, or
if we assume $A^{(2)}{}_{a}{}_{b}$ is symmetric, we solve
for: \begin{equation} A^{(2)}{}_{a}{}_{b} =
\frac{1}{2}\left(-{\left[S{}_{;}{}_{a}\right]}{}_{;}{}_{b} -
  {\left[S{}_{;}{}_{b}\right]}{}_{;}{}_{a} +
  {\left[S\right]}{}_{;}{}_{a}{}_{b} + {\left[S{}_{;}{}_{a}{}_{b}\right]}
\right) \end{equation} We write this as \begin{equation} A^{(2)}{}_{a}{}_{b} \symmeq
-{\left[S{}_{;}{}_{a}\right]}{}_{;}{}_{b} +
  {\frac{{\left[S\right]}{}_{;}{}_{a}{}_{b}}{2}} +
  {\frac{{\left[S{}_{;}{}_{a}{}_{b}\right]}}{2}}
\end{equation} where we use the standard notation $\symmeq$ to denote
equality upon symmetrization. In terms of symbolic processing,
this is implemented by taking each term of a
tensorial expression and putting all free indices in lexigraphic
order. We also define rules that set to zero any Riemann
curvature tensor $R{}_{a}{}_{b}{}_{c}{}_{d}$ when either the
first or second pair of indices are free. For example,
consider when we take three covariant derivatives: \begin{eqnarray}
{\left[S{}_{;}{}_{a}{}_{b}{}_{c}\right]} &=&
A^{(3)}{}_{a}{}_{b}{}_{c} + A^{(3)}{}_{a}{}_{c}{}_{b} +
  A^{(3)}{}_{b}{}_{a}{}_{c} + A^{(3)}{}_{b}{}_{c}{}_{a} +
  A^{(3)}{}_{c}{}_{a}{}_{b} + A^{(3)}{}_{c}{}_{b}{}_{a} \cr
&&        + A^{(2)}{}_{a}{}_{b}{}_{;}{}_{c} +
A^{(2)}{}_{a}{}_{c}{}_{;}{}_{b} +
  A^{(2)}{}_{b}{}_{a}{}_{;}{}_{c} + A^{(2)}{}_{b}{}_{c}{}_{;}{}_{a} +
  A^{(2)}{}_{c}{}_{a}{}_{;}{}_{b} + A^{(2)}{}_{c}{}_{b}{}_{;}{}_{a} \cr
&&        + A^{(1)}{}_{a}{}_{;}{}_{b}{}_{c} +
A^{(1)}{}_{b}{}_{;}{}_{a}{}_{c} +
  A^{(1)}{}_{c}{}_{;}{}_{a}{}_{b} +
  {\frac{A^{(1)}{}_{p}\,R{}_{a}{}_{b}{}_{c}{}^{p}}{3}} +
  {\frac{A^{(1)}{}_{p}\,R{}_{a}{}_{c}{}_{b}{}^{p}}{3}} \cr
&&        + A^{(0)}{}_{;}{}_{a}{}_{b}{}_{c} \end{eqnarray} Now putting all
free indices in lexigraphic order and
then using the above rules for the Riemann curvature
tensor gives 
\begin{eqnarray} {\left[S{}_{;}{}_{a}{}_{b}{}_{c}\right]} &\symmeq&
        6\,A^{(3)}{}_{a}{}_{b}{}_{c} + 6\,A^{(2)}{}_{a}{}_{b}{}_{;}{}_{c} +
  3\,A^{(1)}{}_{a}{}_{;}{}_{b}{}_{c} + A^{(0)}{}_{;}{}_{a}{}_{b}{}_{c} +
  {\frac{2\,A^{(1)}{}_{p}\,R{}_{a}{}_{b}{}_{c}{}^{p}}{3}} \cr
&\symmeq&
        6\,A^{(3)}{}_{a}{}_{b}{}_{c} + 6\,A^{(2)}{}_{a}{}_{b}{}_{;}{}_{c} +
  3\,A^{(1)}{}_{a}{}_{;}{}_{b}{}_{c} + A^{(0)}{}_{;}{}_{a}{}_{b}{}_{c}
\end{eqnarray} This can now be solved for $A^{(3)}{}_{a}{}_{b}{}_{c}$ and
the previously determined results for
$A^{(0)}$, $A^{(1)}{}_{a}$ and $A^{(2)}{}_{a}{}_{b}$
used to get
$A^{(3)}{}_{a}{}_{b}{}_{c}$ solely in terms of the coincident
limit of up to three covariant derivatives acting on $S$. Nothing
new is encountered when determining the rest of the expansion
tensors. We now give the results for up through
$A^{(8)}{}_{a}{}_{b}{}_{c}{}_{d}{}_{e}{}_{f}{}_{g}{}_{h}$: \begin{subequations}\begin{eqnarray}
A^{(3)}{}_{a}{}_{b}{}_{c} &\symmeq&
          {\frac{1}{6}} {\left[S{}_{;}{}_{a}{}_{b}{}_{c}\right]}
          -{\frac{1}{2}} {\left[S{}_{;}{}_{a}{}_{b}\right]}{}_{;}{}_{c}
        + {\frac{1}{2}} {\left[S{}_{;}{}_{a}\right]}{}_{;}{}_{b}{}_{c}
          -{\frac{1}{6}} {\left[S\right]}{}_{;}{}_{a}{}_{b}{}_{c} \\ \cr
A^{(4)}{}_{a}{}_{b}{}_{c}{}_{d} &\symmeq&
           {\frac{1}{24}} {\left[S{}_{;}{}_{a}{}_{b}{}_{c}{}_{d}\right]}
           -{\frac{1}{6}} {\left[S{}_{;}{}_{a}{}_{b}{}_{c}\right]}{}_{;}{}_{d}
         + {\frac{1}{4}} {\left[S{}_{;}{}_{a}{}_{b}\right]}{}_{;}{}_{c}{}_{d}
           -{\frac{1}{6}} {\left[S{}_{;}{}_{a}\right]}{}_{;}{}_{b}{}_{c}{}_{d}
         + {\frac{1}{24}} {\left[S\right]}{}_{;}{}_{a}{}_{b}{}_{c}{}_{d} \cr &&\\ \cr
A^{(5)}{}_{a}{}_{b}{}_{c}{}_{d}{}_{e} &\symmeq&
           {\frac{1}{120}} {\left[S{}_{;}{}_{a}{}_{b}{}_{c}{}_{d}{}_{e}\right]}
           -{\frac{1}{24}}
{\left[S{}_{;}{}_{a}{}_{b}{}_{c}{}_{d}\right]}{}_{;}{}_{e}
         + {\frac{1}{12}}
{\left[S{}_{;}{}_{a}{}_{b}{}_{c}\right]}{}_{;}{}_{d}{}_{e}
           -{\frac{1}{12}}
{\left[S{}_{;}{}_{a}{}_{b}\right]}{}_{;}{}_{c}{}_{d}{}_{e} \cr
&&         + {\frac{1}{24}}
{\left[S{}_{;}{}_{a}\right]}{}_{;}{}_{b}{}_{c}{}_{d}{}_{e}
           -{\frac{1}{120}}
{\left[S\right]}{}_{;}{}_{a}{}_{b}{}_{c}{}_{d}{}_{e} \\ \cr
A^{(6)}{}_{a}{}_{b}{}_{c}{}_{d}{}_{e}{}_{f} &\symmeq&
           {\frac{1}{720}}
{\left[S{}_{;}{}_{a}{}_{b}{}_{c}{}_{d}{}_{e}{}_{f}\right]}
           -{\frac{1}{120}}
{\left[S{}_{;}{}_{a}{}_{b}{}_{c}{}_{d}{}_{e}\right]}{}_{;}{}_{f}
         + {\frac{1}{48}}
{\left[S{}_{;}{}_{a}{}_{b}{}_{c}{}_{d}\right]}{}_{;}{}_{e}{}_{f}
           -{\frac{1}{36}}
{\left[S{}_{;}{}_{a}{}_{b}{}_{c}\right]}{}_{;}{}_{d}{}_{e}{}_{f}
\cr &&         + {\frac{1}{48}}
{\left[S{}_{;}{}_{a}{}_{b}\right]}{}_{;}{}_{c}{}_{d}{}_{e}{}_{f}
           -{\frac{1}{120}}
{\left[S{}_{;}{}_{a}\right]}{}_{;}{}_{b}{}_{c}{}_{d}{}_{e}{}_{f}
         + {\frac{1}{720}}
{\left[S\right]}{}_{;}{}_{a}{}_{b}{}_{c}{}_{d}{}_{e}{}_{f} \\ \cr
A^{(7)}{}_{a}{}_{b}{}_{c}{}_{d}{}_{e}{}_{f}{}_{g} &\symmeq&
           {\frac{1}{5040}}
{\left[S{}_{;}{}_{a}{}_{b}{}_{c}{}_{d}{}_{e}{}_{f}{}_{g}\right]}
           -{\frac{1}{720}}
{\left[S{}_{;}{}_{a}{}_{b}{}_{c}{}_{d}{}_{e}{}_{f}\right]}{}_{;}{}_{g}
         + {\frac{1}{240}}
{\left[S{}_{;}{}_{a}{}_{b}{}_{c}{}_{d}{}_{e}\right]}{}_{;}{}_{f}{}_{g}
\cr &&           -{\frac{1}{144}}
{\left[S{}_{;}{}_{a}{}_{b}{}_{c}{}_{d}\right]}{}_{;}{}_{e}{}_{f}{}_{g}
         + {\frac{1}{144}}
{\left[S{}_{;}{}_{a}{}_{b}{}_{c}\right]}{}_{;}{}_{d}{}_{e}{}_{f}{}_{g}
           -{\frac{1}{240}}
{\left[S{}_{;}{}_{a}{}_{b}\right]}{}_{;}{}_{c}{}_{d}{}_{e}{}_{f}{}_{g}
\cr &&         + {\frac{1}{720}}
{\left[S{}_{;}{}_{a}\right]}{}_{;}{}_{b}{}_{c}{}_{d}{}_{e}{}_{f}{}_{g}
           -{\frac{1}{5040}}
{\left[S\right]}{}_{;}{}_{a}{}_{b}{}_{c}{}_{d}{}_{e}{}_{f}{}_{g}
\\ \cr A^{(8)}{}_{a}{}_{b}{}_{c}{}_{d}{}_{e}{}_{f}{}_{g}{}_{h}
&\symmeq&
           {\frac{1}{40320}}
{\left[S{}_{;}{}_{a}{}_{b}{}_{c}{}_{d}{}_{e}{}_{f}{}_{g}{}_{h}\right]}
           -{\frac{1}{5040}}
{\left[S{}_{;}{}_{a}{}_{b}{}_{c}{}_{d}{}_{e}{}_{f}{}_{g}\right]}{}_{;}{}_{h}
         + {\frac{1}{1440}}
{\left[S{}_{;}{}_{a}{}_{b}{}_{c}{}_{d}{}_{e}{}_{f}\right]}{}_{;}{}_{g}{}_{h}
\cr &&           -{\frac{1}{720}}
{\left[S{}_{;}{}_{a}{}_{b}{}_{c}{}_{d}{}_{e}\right]}{}_{;}{}_{f}{}_{g}{}_{h}
         + {\frac{1}{576}}
{\left[S{}_{;}{}_{a}{}_{b}{}_{c}{}_{d}\right]}{}_{;}{}_{e}{}_{f}{}_{g}{}_{h}
           -{\frac{1}{720}}
{\left[S{}_{;}{}_{a}{}_{b}{}_{c}\right]}{}_{;}{}_{d}{}_{e}{}_{f}{}_{g}{}_{h}
\cr &&         + {\frac{1}{1440}}
{\left[S{}_{;}{}_{a}{}_{b}\right]}{}_{;}{}_{c}{}_{d}{}_{e}{}_{f}{}_{g}{}_{h}
           -{\frac{1}{5040}}
{\left[S{}_{;}{}_{a}\right]}{}_{;}{}_{b}{}_{c}{}_{d}{}_{e}{}_{f}{}_{g}{}_{h}
         + {\frac{1}{40320}}
{\left[S\right]}{}_{;}{}_{a}{}_{b}{}_{c}{}_{d}{}_{e}{}_{f}{}_{g}{}_{h}
\cr && \end{eqnarray}\end{subequations}

These relations simplify considerable if the scalar $S$ is
symmetric, $S(x,y) = S(y,x)$, for the
symmetrized odd derivatives are determined by the even
derivatives. For one derivative, \begin{equation} S(x,y)_{;a'} = S(y,x)_{;a'}
\quad \Rightarrow \quad {\left[S{}_{;}{}_{a'}\right]} =
{\left[S{}_{;}{}_{a}\right]}, \end{equation} and applying Synge's theorem to
the left-hand side above yields \begin{equation}
{\left[S\right]}{}_{;}{}_{a} - {\left[S{}_{;}{}_{a}\right]} =
{\left[S{}_{;}{}_{a}\right]} \quad \Rightarrow \quad
{\left[S{}_{;}{}_{a}\right]} =
{\frac{{\left[S\right]}{}_{;}{}_{a}}{2}} \label{Apx-odd1rule} \end{equation}
For three derivatives, we have \begin{equation}
{\left[S{}_{;}{}_{a'}{}_{b'}{}_{c'}\right]} =
{\left[S{}_{;}{}_{a}{}_{b}{}_{c}\right]} \end{equation} Once again, using
Synge's theorem and (\ref{Apx-odd1rule}): \begin{equation}
{\left[S{}_{;}{}_{a}{}_{b}\right]}{}_{;}{}_{c} +
  {\left[S{}_{;}{}_{a}{}_{c}\right]}{}_{;}{}_{b} +
  {\left[S{}_{;}{}_{b}{}_{c}\right]}{}_{;}{}_{a} -
  {\frac{{\left[S\right]}{}_{;}{}_{a}{}_{c}{}_{b}}{2}} -
  {\left[S{}_{;}{}_{b}{}_{c}{}_{a}\right]} =
{\left[S{}_{;}{}_{a}{}_{b}{}_{c}\right]} \end{equation} It follows from
this that
\begin{subequations}
\begin{equation} 4\,{\left[S{}_{;}{}_{(}\!{}_{a}{}_{b}{}_{c}\!{}_{)}\right]} =
6\,{\left[S{}_{;}{}_{(}\!{}_{a}{}_{b}\right]}{}_{;}{}_{c}\!{}_{)}
-
  {\left[S\right]}{}_{;}{}_{(}\!{}_{a}{}_{b}{}_{c}\!{}_{)}
\label{apx-endpoint-odd3} \end{equation} The results for five and seven
derivatives are \begin{eqnarray}
2\,{\left[S{}_{;}{}_{(}\!{}_{a}{}_{b}{}_{c}{}_{d}{}_{e}\!{}_{)}\right]}
&=&
5\,{\left[S{}_{;}{}_{(}\!{}_{a}{}_{b}{}_{c}{}_{d}\right]}{}_{;}
    {}_{e}\!{}_{)} - 5\,{\left[S{}_{;}{}_{(}\!{}_{a}{}_{b}\right]}{}_{;}{}_{c}
    {}_{d}{}_{e}\!{}_{)} + {\left[S\right]}{}_{;}{}_{(}\!{}_{a}{}_{b}{}_{c}
   {}_{d}{}_{e}\!{}_{)} \\
8\,{\left[S{}_{;}{}_{(}\!{}_{a}{}_{b}{}_{c}{}_{d}{}_{e}{}_{f}{}_{g}\!{}_{)}
   \right]} &=&
28\,{\left[S{}_{;}{}_{(}\!{}_{a}{}_{b}{}_{c}{}_{d}{}_{e}{}_{f}\right]}{}_{;}
    {}_{g}\!{}_{)} - 70\,{\left[S{}_{;}{}_{(}\!{}_{a}{}_{b}{}_{c}{}_{d}
     \right]}{}_{;}{}_{e}{}_{f}{}_{g}\!{}_{)} +
  84\,{\left[S{}_{;}{}_{(}\!{}_{a}{}_{b}\right]}{}_{;}{}_{c}{}_{d}{}_{e}{}_{f}
    {}_{g}\!{}_{)} \cr
&&
-17\,{\left[S\right]}{}_{;}{}_{(}\!{}_{a}{}_{b}{}_{c}{}_{d}{}_{e}{}_{f}
   {}_{g}\!{}_{)}
\end{eqnarray}
\end{subequations}
With these results, the equations for the even expansion tensors
for a symmetric function simplify: \begin{subequations}\begin{eqnarray}
A^{(0)} &=& {\left[S\right]} \\
2 ! A^{(2)}{}_{a}{}_{b} &=&
  {\left[S{}_{;}{}_{a}{}_{b}\right]} \\
4 ! A^{(4)}{}_{a}{}_{b}{}_{c}{}_{d} &\symmeq&
  {\left[S{}_{;}{}_{a}{}_{b}{}_{c}{}_{d}\right]} \\
6 ! A^{(6)}{}_{a}{}_{b}{}_{c}{}_{d}{}_{e}{}_{f} &\symmeq&
  {\left[S{}_{;}{}_{a}{}_{b}{}_{c}{}_{d}{}_{e}{}_{f}\right]} \\
8 ! A^{(8)}{}_{a}{}_{b}{}_{c}{}_{d}{}_{e}{}_{f}{}_{g}{}_{h}
&\symmeq&
  {\left[S{}_{;}{}_{a}{}_{b}{}_{c}{}_{d}{}_{e}{}_{f}{}_{g}{}_{h}\right]}
\label{Apx-evenArules} \end{eqnarray}\end{subequations} while the odd expansion tensors
are given in terms of the even tensors: \begin{subequations}\begin{eqnarray} 2 ! A^{(1)}{}_{a}
&\symmeq&
  -A^{(0)}{}_{;}{}_{a} \\
4 ! A^{(3)}{}_{a}{}_{b}{}_{c} &\symmeq&
  -12\,A^{(2)}{}_{a}{}_{b}{}_{;}{}_{c} + A^{(0)}{}_{;}{}_{a}{}_{b}{}_{c} \\
6 ! A^{(5)}{}_{a}{}_{b}{}_{c}{}_{d}{}_{e} &\symmeq&
  -360\,A^{(4)}{}_{a}{}_{b}{}_{c}{}_{d}{}_{;}{}_{e} +
  30\,A^{(2)}{}_{a}{}_{b}{}_{;}{}_{c}{}_{d}{}_{e} -
  3\,A^{(0)}{}_{;}{}_{a}{}_{b}{}_{c}{}_{d}{}_{e} \\
8 ! A^{(7)}{}_{a}{}_{b}{}_{c}{}_{d}{}_{e}{}_{f}{}_{g} &\symmeq&
  -20160\,A^{(6)}{}_{a}{}_{b}{}_{c}{}_{d}{}_{e}{}_{f}{}_{;}{}_{g} +
  1680\,A^{(4)}{}_{a}{}_{b}{}_{c}{}_{d}{}_{;}{}_{e}{}_{f}{}_{g} -
  168\,A^{(2)}{}_{a}{}_{b}{}_{;}{}_{c}{}_{d}{}_{e}{}_{f}{}_{g} \cr
  && + 17\,A^{(0)}{}_{;}{}_{a}{}_{b}{}_{c}{}_{d}{}_{e}{}_{f}{}_{g}
\label{Apx-oddArules} \end{eqnarray}\end{subequations}

Very often, we need the covariant derivative of a series expansion
(\ref{Apx-series1}): \begin{eqnarray} S{}_{;}{}_{a} &=&
        A^{(0)}{}_{;}{}_{a} + \sigma{}^{;}{}^{p}\,A^{(1)}{}_{p}{}_{;}{}_{a} +
  A^{(1)}{}_{p}\,\sigma{}_{;}{}_{a}{}^{p} \cr
        &&+
\sigma{}^{;}{}^{p}\,\sigma{}^{;}{}^{q}\,A^{(2)}{}_{p}{}_{q}{}_{;}{}_{a}
+
  A^{(2)}{}_{p}{}_{q}\,\sigma{}^{;}{}^{q}\,\sigma{}_{;}{}_{a}{}^{p} +
  A^{(2)}{}_{p}{}_{q}\,\sigma{}^{;}{}^{p}\,\sigma{}_{;}{}_{a}{}^{q} \cdots
\end{eqnarray} If we replace $\sigma{}_{;}{}_{a}{}_{b}$ with its series
expansion, then we readily have the series expansion of
$S{}_{;}{}_{a}$. We can get this via the above relations by
merely replacing $S$ with $\sigma{}_{;}{}_{a}{}_{b}$. In
particular, we want the expansion \begin{eqnarray} \sigma{}_{;}{}_{a}{}_{b}
&=& B^{(0)}{}_{a}{}_{b} +
B^{(1)}{}_{a}{}_{b}{}_{p}\,\sigma{}^{;}{}^{p} +
  B^{(2)}{}_{a}{}_{b}{}_{p}{}_{q}\,\sigma{}^{;}{}^{p}\,\sigma{}^{;}{}^{q} +
  B^{(3)}{}_{a}{}_{b}{}_{p}{}_{q}{}_{r}\,\sigma{}^{;}{}^{p}\,
   \sigma{}^{;}{}^{q}\,\sigma{}^{;}{}^{r} \cr
&&+
B^{(4)}{}_{a}{}_{b}{}_{p}{}_{q}{}_{r}{}_{s}\,\sigma{}^{;}{}^{p}\,
   \sigma{}^{;}{}^{q}\,\sigma{}^{;}{}^{r}\,\sigma{}^{;}{}^{s} +
  B^{(5)}{}_{a}{}_{b}{}_{p}{}_{q}{}_{r}{}_{s}{}_{t}\,\sigma{}^{;}{}^{p}\,
   \sigma{}^{;}{}^{q}\,\sigma{}^{;}{}^{r}\,\sigma{}^{;}{}^{s}\,
   \sigma{}^{;}{}^{t} \cr
&&+
B^{(6)}{}_{a}{}_{b}{}_{p}{}_{q}{}_{r}{}_{s}{}_{t}{}_{u}\,\sigma{}^{;}{}^{p}\,
  \sigma{}^{;}{}^{q}\,\sigma{}^{;}{}^{r}\,\sigma{}^{;}{}^{s}\,
  \sigma{}^{;}{}^{t}\,\sigma{}^{;}{}^{u}
\label{Apx-cd2sigmaseries} \end{eqnarray} We immediately have \begin{equation}
B^{(0)}{}_{a}{}_{b} = {\left[\sigma{}_{;}{}_{a}{}_{b}\right]}
 = g{}_{a}{}_{b}
\end{equation} and \begin{equation} B^{(1)}{}_{a}{}_{b}{}_{c} =
-{\left[\sigma{}_{;}{}_{a}{}_{b}\right]}{}_{;}{}_{c} +
  {\left[\sigma{}_{;}{}_{a}{}_{b}{}_{c}\right]}
 =  0
\end{equation} For the second order term: \begin{eqnarray}
B^{(2)}{}_{a}{}_{b}{}_{c}{}_{d} &=&
-{\left[\sigma{}_{;}{}_{a}{}_{b}{}_{c}\right]}{}_{;}{}_{d} +
  {\frac{{\left[\sigma{}_{;}{}_{a}{}_{b}\right]}{}_{;}{}_{c}{}_{d}}{2}} +
  {\frac{{\left[\sigma{}_{;}{}_{a}{}_{b}{}_{c}{}_{d}\right]}}{2}}\cr
 &=& -\frac{1}{6}\left(
        R{}_{a}{}_{c}{}_{b}{}_{d} + R{}_{a}{}_{d}{}_{b}{}_{c}
        \right)
\end{eqnarray} Now we have to be careful about how we carry out the
symmetrization: it is only the indices $c,d$ that are contracted
over in the series (\ref{Apx-cd2sigmaseries}). So it is only the
free indices {\em other than} $a$ and $b$ in this and the following
that we put in lexographic order (our routine for ordering free
indices can be given a list of indices to exclude from ordering):
\begin{equation} B^{(2)}{}_{a}{}_{b}{}_{c}{}_{d} \symmeq'
-{\frac{R{}_{a}{}_{c}{}_{b}{}_{d}}{3}} \end{equation} Equality upon
symmetrization of all indices but $a,b$ is denoted by $\symmeq'$.
The rest of the expansion tensors are computed in the same way;
the results are \begin{subequations}\begin{eqnarray}
B^{(3)}{}_{a}{}_{b}{}_{c}{}_{d}{}_{e} &\symmeq'&
  {\frac{R{}_{a}{}_{c}{}_{b}{}_{d}{}_{;}{}_{e}}{12}} \\
B^{(4)}{}_{a}{}_{b}{}_{c}{}_{d}{}_{e}{}_{f} &\symmeq'&
  {\frac{1}{180}} \left(
   -3\,R{}_{a}{}_{c}{}_{b}{}_{d}{}_{;}{}_{e}{}_{f} +
  4\,R{}_{p}{}_{c}{}_{a}{}_{d}\,R{}_{b}{}_{e}{}_{f}{}^{p}
  \right) \\
B^{(5)}{}_{a}{}_{b}{}_{c}{}_{d}{}_{e}{}_{f}{}_{g} &\symmeq'&
  {\frac{1}{360}} \left(
   R{}_{a}{}_{c}{}_{b}{}_{d}{}_{;}{}_{e}{}_{f}{}_{g} -
  3\,R{}_{p}{}_{d}{}_{b}{}_{c}{}_{;}{}_{e}\,R{}_{a}{}_{g}{}_{f}{}^{p} -
  3\,R{}_{p}{}_{d}{}_{a}{}_{c}{}_{;}{}_{e}\,R{}_{b}{}_{g}{}_{f}{}^{p}
  \right) \\
B^{(6)}{}_{a}{}_{b}{}_{c}{}_{d}{}_{e}{}_{f}{}_{g}{}_{h} &\symmeq'&
  {\frac{1}{15120}} \left(
                 -219\,R{}_{p}{}_{d}{}_{a}{}_{c}{}_{;}{}_{e}\,
   R{}_{b}{}_{f}{}_{g}{}^{p}{}_{;}{}_{h} -
  6\,R{}_{a}{}_{c}{}_{b}{}_{d}{}_{;}{}_{e}{}_{f}{}_{g}{}_{h} \right. \cr
&&        \hspace{10mm}
-114\,R{}_{p}{}_{d}{}_{b}{}_{c}{}_{;}{}_{e}{}_{f}\,
   R{}_{a}{}_{g}{}_{h}{}^{p} - 114\,
   R{}_{p}{}_{d}{}_{a}{}_{c}{}_{;}{}_{e}{}_{f}\,R{}_{b}{}_{g}{}_{h}{}^{p} \cr
&&        \hspace{10mm}\left.+
24\,R{}_{p}{}_{e}{}_{a}{}_{f}\,R{}_{q}{}_{h}{}_{b}{}_{g}\,
   R{}_{c}{}^{p}{}_{d}{}^{q} + 8\,R{}_{p}{}_{e}{}_{a}{}_{f}\,
   R{}_{q}{}_{h}{}_{b}{}_{g}\,R{}_{c}{}^{q}{}_{d}{}^{p}
  \right)
\label{Apx-cd2sigmacoef} \end{eqnarray}\end{subequations}

%%%%%%%%%%%%%%%%%%%%%%%%%%%%%%%%%%%%%%%%%%%%%%%%
\section{VanVleck-Morette Determinant}
\label{apx-VanD}

Other than the world function, the other main geometric object we
need is the VanVlette-Morette determinant, defined as \begin{equation} D(x,y)
\equiv - \det\left(-\sigma_{;ab'}\right). \end{equation} In the context of
Green function, what appears is \begin{equation} \VanD(x,y) =
\left(\frac{D(x,y)}{\sqrt{g(x)g(y)}}\right)^\frac{1}{2} \end{equation}
which we focus on. Using $2\sigma=\sigma^{;p}\sigma_{;p}$ the
VanVlette-Morette determinant is seen to satisfy \begin{equation}
D^{-1}\left(D \sigma^{;p}\right)_{;p} = 4 \quad \Rightarrow \quad
\VanD\left(4-\square\sigma\right) - 2 \VanD_{\,\,;p} \sigma^{;p} = 0
\label{Apx-VanD1} \end{equation} along with \begin{equation} \left[ D \right ] = g(x)
\quad \Rightarrow \quad \left[ \VanD \right ] =1 \end{equation} from which
we readily get \begin{equation} \left[ \VanD_{\,\,;a} \right ] = 0. \end{equation}

We could at this point proceed as we did with $\sigma$ to
determine the coincident limit expression of covariant
derivatives of $\VanD$. But what we need is the end point
expansion of $\VanD(x,y)$ to sixth order in $\sigma^a$. With this
in mind, we set out to directly determine the series. We start by
assuming the expansion \begin{eqnarray} \VanD &=& 1
 + \Delta^{(2)}_{pq}    \sigma^p \sigma^q
 + \Delta^{(3)}_{pqr}   \sigma^p \sigma^q \sigma^r
 + \Delta^{(4)}_{pqrs}  \sigma^p \sigma^q \sigma^r \sigma^s
 + \Delta^{(5)}_{pqrst} \sigma^p \sigma^q \sigma^r \sigma^s \sigma^t \cr
&&\hspace{2cm}
 + \Delta^{(6)}_{pqrstu}\sigma^p \sigma^q \sigma^r \sigma^s \sigma^t \sigma^u
\label{Apx-VanDseries} \end{eqnarray} and substitute back into
(\ref{Apx-VanD1}). We use (\ref{Apx-cd2sigmaseries}) and
$\sigma^a = \left(2\sigma\right)^\frac{1}{2} p^a$. The expansion
tensors $\Delta^{(n)}_{a_1\cdots a_n}$ are determined by
collecting terms according to their order in $\sigma$ and setting
to zero.

From (\ref{Apx-cd2sigmaseries}) and (\ref{Apx-cd2sigmacoef}), we
have the series expansion \begin{eqnarray} \square \sigma &=&  4
-\frac{2}{3}  \sigma\,p{}^{p}\,p{}^{q}\,R{}_{p}{}_{q}
+\frac{\sqrt{2}}{6}
{{\sigma}^{{\frac{3}{2}}}}\,R{}_{p}{}_{q}{}_{;}{}_{r}\,p{}^{p}\,p{}^{q}\,
  p{}^{r}
\cr && -\frac{1}{45}
{{\sigma}^2}\,p{}^{p}\,p{}^{q}\,p{}^{r}\,p{}^{s}\,
  \left( 3\,R{}_{p}{}_{q}{}_{;}{}_{r}{}_{s} +
    4\,R{}_{p}{}_{t}{}_{q}{}_{u}\,R{}_{r}{}^{u}{}_{s}{}^{t} \right)
\cr && +\frac{\sqrt{2}}{90}
    {{\sigma}^{{\frac{5}{2}}}}\,p{}^{p}\,p{}^{q}\,p{}^{r}\,p{}^{s}\,p{}^{t}\,
  \left( R{}_{p}{}_{q}{}_{;}{}_{r}{}_{s}{}_{t} +
    6\,R{}_{p}{}_{v}{}_{q}{}_{u}{}_{;}{}_{r}\,R{}_{s}{}^{v}{}_{t}{}^{u}
    \right)
\cr && +\frac{1}{1890}
  {{\sigma}^3}\,p{}^{p}\,p{}^{q}\,p{}^{r}\,p{}^{s}\,p{}^{t}\,p{}^{u} \left(
     219\,R{}_{p}{}_{v}{}_{q}{}_{w}{}_{;}{}_{r}\,
   R{}_{s}{}^{v}{}_{t}{}^{w}{}_{;}{}_{u} -
  6\,R{}_{p}{}_{q}{}_{;}{}_{r}{}_{s}{}_{t}{}_{u}
\right. \cr && \hspace{15mm}\left. +
     24\,R{}_{p}{}_{v}{}_{q}{}_{w}\,R{}_{r}{}^{w}{}_{s}{}_{x}\,
   R{}_{t}{}^{v}{}_{u}{}^{x} + 228\,
   R{}_{p}{}_{w}{}_{q}{}_{v}{}_{;}{}_{r}{}_{s}\,R{}_{t}{}^{w}{}_{u}{}^{v}
\right. \cr && \hspace{15mm}\left. +
     8\,R{}_{p}{}_{v}{}_{q}{}_{w}\,R{}_{r}{}^{w}{}_{s}{}_{x}\,
  R{}_{t}{}^{x}{}_{u}{}^{v}
  \right)
\end{eqnarray} We have made the split into $\sigma$ and $p^a$ so we can
readily carry out the needed multiplication in the series
(\ref{Apx-VanDseries}), once we put it into the same form. The
last piece we need is $\VanD_{\,;a}$. This is obtained by
differentiating (\ref{Apx-VanDseries}) and then substituting
(\ref{Apx-cd2sigmaseries}) for the $\sigma_{;ab}$ terms that
arise. Once this is done, we have all the terms for the series
expansion of (\ref{Apx-VanD1}).
The order $\sigma$ term: \begin{equation}
{\frac{2\,\sigma\,p{}^{p}\,p{}^{q}}{3}} \left(
        12\,\Delta^{(2)}{}_{p}{}_{q} - R{}_{p}{}_{q} \right) = 0
\quad\Rightarrow\quad \Delta^{(2)}{}_{a}{}_{b} =
{\frac{R{}_{a}{}_{b}}{12}} \label{Apx-D2rule} \end{equation}
The order $\sigma^\frac{3}{2}$ term starts off as \begin{eqnarray} \lefteqn{
{\frac{{{\sigma}^{{\frac{3}{2}}}}\,p{}^{p}\,p{}^{q}\,p{}^{r}}{3\,{\sqrt{2}}}}
\left(
    24\,\Delta^{(2)}{}_{p}{}_{q}{}_{;}{}_{r} + R{}_{p}{}_{q}{}_{;}{}_{r} +
  72\,\Delta^{(3)}{}_{p}{}_{q}{}_{r}
\right) = 0 }\cr &&\quad\Rightarrow\quad
\Delta^{(3)}{}_{a}{}_{b}{}_{c} \symmeq -\frac{1}{72} \left(
24\,\Delta^{(2)}{}_{a}{}_{b}{}_{;}{}_{c} +
R{}_{a}{}_{b}{}_{;}{}_{c} \right). \end{eqnarray} Using (\ref{Apx-D2rule}) shows
\begin{equation} \Delta^{(3)}{}_{a}{}_{b}{}_{c} \symmeq -
{\frac{R{}_{a}{}_{b}{}_{;}{}_{c}}{24}} \label{Apx-D3rule} \end{equation}
For the order $\sigma^2$ term, we find \begin{eqnarray}
\Delta^{(4)}{}_{a}{}_{b}{}_{c}{}_{d} &\symmeq&
\frac{1}{1440}\left(
-360\,\Delta^{(3)}{}_{a}{}_{b}{}_{c}{}_{;}{}_{d} +
  3\,R{}_{a}{}_{b}{}_{;}{}_{c}{}_{d} +
  60\,\Delta^{(2)}{}_{a}{}_{b}\,R{}_{c}{}_{d} +
  120\,\Delta^{(2)}{}_{p}{}_{a}\,R{}_{b}{}_{c}{}_{d}{}^{p} \right. \cr
&& \hspace{10mm} \left. +
120\,\Delta^{(2)}{}_{a}{}_{p}\,R{}_{b}{}^{p}{}_{c}{}_{d} +
  4\,R{}_{p}{}_{a}{}_{q}{}_{b}\,R{}_{c}{}^{q}{}_{d}{}^{p} \right)
\end{eqnarray} The fourth and fifth terms above vanish since we only need
equality up to symmetrization. Using (\ref{Apx-D2rule}) and
(\ref{Apx-D3rule}), the final form for this term becomes
\begin{subequations}
\begin{equation} \Delta^{(4)}{}_{a}{}_{b}{}_{c}{}_{d} \symmeq
\frac{1}{1440}\left( 18\,R{}_{a}{}_{b}{}_{;}{}_{c}{}_{d} +
5\,R{}_{a}{}_{b}\,R{}_{c}{}_{d} +
  4\,R{}_{p}{}_{a}{}_{q}{}_{b}\,R{}_{c}{}^{q}{}_{d}{}^{p} \right)
\label{Apx-D4rule} \end{equation} The last two coefficients are determined
in exactly the same manner. The results are: \begin{eqnarray}
\Delta^{(5)}{}_{a}{}_{b}{}_{c}{}_{d}{}_{e} &\symmeq&
-\frac{1}{1440} \left( 4\,R{}_{a}{}_{b}{}_{;}{}_{c}{}_{d}{}_{e} +
  5\,R{}_{a}{}_{b}{}_{;}{}_{c}\,R{}_{d}{}_{e} +
  4\,R{}_{p}{}_{a}{}_{q}{}_{b}{}_{;}{}_{c}\,R{}_{d}{}^{q}{}_{e}{}^{p}
\right) \\
\Delta^{(6)}{}_{a}{}_{b}{}_{c}{}_{d}{}_{e}{}_{f} &\symmeq&
{\frac{1}{362880}} \left(
  315\,R{}_{a}{}_{b}{}_{;}{}_{c}\,R{}_{d}{}_{e}{}_{;}{}_{f} +
  180\,R{}_{a}{}_{b}{}_{;}{}_{c}{}_{d}{}_{e}{}_{f} +
  378\,R{}_{a}{}_{b}{}_{;}{}_{c}{}_{d}\,R{}_{e}{}_{f}
\right. \cr &&\hspace{10mm}   \left.
  35\,R{}_{a}{}_{b}\,R{}_{c}{}_{d}\,R{}_{e}{}_{f} +
84\,R{}_{a}{}_{b}\,R{}_{p}{}_{c}{}_{q}{}_{d}\,R{}_{e}{}^{q}{}_{f}{}^{p}
-270\,R{}_{p}{}_{a}{}_{q}{}_{b}{}_{;}{}_{c}\,
  R{}_{d}{}^{q}{}_{e}{}^{p}{}_{;}{}_{f}
\right. \cr &&\hspace{10mm}   \left.
-288\,R{}_{p}{}_{a}{}_{q}{}_{b}{}_{;}{}_{c}{}_{d}\,
   R{}_{e}{}^{q}{}_{f}{}^{p} + 64\,R{}_{p}{}_{a}{}_{q}{}_{b}\,
   R{}_{r}{}_{d}{}_{c}{}^{q}\,R{}_{e}{}^{r}{}_{f}{}^{p} \right)
\end{eqnarray}
\end{subequations}

%%%%%%%%%%%%%%%%%%%%%%%%%%%%%%%%%%%%%%%%%%%%%%
\section{Series expansion for $\Delta\tau$}
\label{apx-dtseries}

In this section we determine the expansion \begin{equation} {{\Delta\tau}^2} =
\sigma{}^{;}{}^{p}\,\sigma{}^{;}{}^{q}\,\delta\tau^{(2)}{}_{p}{}_{q}
+
  \sigma{}^{;}{}^{p}\,\sigma{}^{;}{}^{q}\,\sigma{}^{;}{}^{r}\,
   \delta\tau^{(3)}{}_{p}{}_{q}{}_{r} +
  \sigma{}^{;}{}^{p}\,\sigma{}^{;}{}^{q}\,\sigma{}^{;}{}^{r}\,
   \sigma{}^{;}{}^{s}\,\delta\tau^{(4)}{}_{p}{}_{q}{}_{r}{}_{s}.
\end{equation} Since this is the expansion of a symmetric function, the
expansion tensors are related the coincident limit of the
covariant derivatives of $\Delta\tau^2$ via
(\ref{Apx-evenArules}) and (\ref{Apx-oddArules}). We also use
$\left[ \Delta\tau \right] = 0$, justifying the series expansion
starts at order two in $\sigma^a$. The expansion tensors are \begin{subequations}\begin{eqnarray}
\delta\tau^{(2)}{}_{a}{}_{b} &\symmeq&
  {\left[\Delta\tau{}_{;}{}_{a}\right]}\,{\left[\Delta\tau{}_{;}{}_{b}\right]}
\\
\delta\tau^{(3)}{}_{a}{}_{b}{}_{c} &\symmeq&
  -\left( {\left[\Delta\tau{}_{;}{}_{a}\right]}{}_{;}{}_{b}\,
    {\left[\Delta\tau{}_{;}{}_{c}\right]} \right)
\\
\delta\tau^{(4)}{}_{a}{}_{b}{}_{c}{}_{d} &\symmeq&
  {\frac{{\left[\Delta\tau{}_{;}{}_{a}{}_{b}\right]}\,
      {\left[\Delta\tau{}_{;}{}_{c}{}_{d}\right]}}{4}} +
  {\frac{{\left[\Delta\tau{}_{;}{}_{a}\right]}\,
      {\left[\Delta\tau{}_{;}{}_{b}{}_{c}{}_{d}\right]}}{3}}
\label{apx-dt1} \end{eqnarray}\end{subequations}

To evaluate the covariant derivatives, we start with
$\Delta\tau{}_{;}{}_{a} = \Delta\tau{}_{,}{}_{a} =
{\delta}{}_{a}{}^{\tau}$. Also: $\Delta\tau_{,ab} = 0$ and
$\left[ \Delta\tau{}_{;}{}_{a} \right] = \Delta\tau{}_{;}{}_{a}
\Rightarrow {\left[\Delta\tau{}_{;}{}_{a}\right]}{}_{;}{}_{b} =
\Delta\tau{}_{;}{}_{a}{}_{b}$.

Turning to the computation of two covariant derivatives: \begin{eqnarray}
\Delta\tau{}_{;}{}_{a}{}_{b} &=&
  \Delta\tau{}_{;}{}_{a}{}_{,}{}_{b} -
\Gamma{}^{p}{}_{a}{}_{b}\,\Delta\tau{}_{;}{}_{p} \cr &=&
  \Delta\tau{}_{;}{}_{a}{}_{,}{}_{b}
                                          -
\Gamma{}^{p}{}_{a}{}_{b}\,\Delta\tau{}_{;}{}_{p} \cr &=&
 -  \Gamma{}^{p}{}_{a}{}_{b}\,{\delta}{}_{p}{}^{\tau}
\cr &=&
    -\Gamma{}^{\tau}{}_{a}{}_{b}
\end{eqnarray} and three covariant derivatives: \begin{equation}
\Delta\tau{}_{;}{}_{a}{}_{b}{}_{c} =
-\Gamma{}^{\tau}{}_{a}{}_{b}{}_{;}{}_{c} \end{equation} Using these
results, the expansion tensors (\ref{apx-dt1}) become 
\begin{subequations}
\label{Apx-dtrules}
\begin{eqnarray}
\delta\tau^{(2)}{}_{a}{}_{b} &=&
        {\delta}{}_{a}{}^{\tau}\,{\delta}{}_{b}{}^{\tau} \\
 \delta\tau^{(3)}{}_{a}{}_{b}{}_{c} &=&
        \Gamma{}^{\tau}{}_{a}{}_{b}\,{\delta}{}_{c}{}^{\tau} \\ 
\delta\tau^{(4)}{}_{a}{}_{b}{}_{c}{}_{d} &=&
\frac{1}{4}
\Gamma{}^{\tau}{}_{a}{}_{b}\,\Gamma{}^{\tau}{}_{c}{}_{d} -
\frac{1}{3}
\Gamma{}^{\tau}{}_{a}{}_{b}{}_{;}{}_{c}\,{\delta}{}_{d}{}^{\tau}
\end{eqnarray}\end{subequations}

%%%%%%%%%%%%%%%%%%%%%%%%%%%%%%%%%%%%%%%%%%%%%%
\section{Series expansion of Hadamard Form}
\label{sec-Had-form}

For the regularization of the coincident limit of the noise kernel,
we need the Hadamard form
\begin{equation}
S(x,y) = \frac{1}{(4\pi)^2}\left(
  \frac{2\VanD}{\sigma} +
  \left(v_0 + \sigma v_1 + \sigma^2 v_2\right)\log\sigma +
  \left(\sigma w_1 + \sigma^2 w_2\right)
\right)
\end{equation}
 to fourth order in $\sigma^a$. We review the standard techniques
for finding these expansions and present the results we use.
The functions $v_n(x,y)$ and $w_n(x,y)$, $ n \ge 1$ are determined by
demanding
\begin{equation}
\left(\square - R/6\right)S(x,y) = 0.
\label{Apx-Had-eqn}
\end{equation}
The arbitrary function $w_0(x,y)$ is assumed to vanish. Working to
fourth order, we proceed in the now familiar pattern of expanding
each of the function in a series expansion and then solve for the
expansion tensors by putting the expansion in the equations derived from
(\ref{Apx-Had-eqn}). Using the differential operators
\begin{subequations}\begin{eqnarray}
H_n &=& \sigma^{;p} \nabla_p + \left( (n - 1 +
            \frac{1}{2} \left(\square\sigma)\right) \right)\\
K &=& \square  - \frac{R}{6}
\end{eqnarray}\end{subequations}
we need to solve
\begin{subequations}\begin{eqnarray}
v_0(x,y) &=& v^{(0)}_0 + \sigma^p v^{(1)}_{0p}
 + \sigma^p \sigma^q v^{(2)}_{0pq}
 + \sigma^p \sigma^q \sigma^r v^{(3)}_{0pqr}
 + \sigma^p \sigma^q \sigma^r \sigma^s v^{(4)}_{0pqrs} \\
H_0 v_0 &=& - K \VanD
\end{eqnarray}\end{subequations}
\begin{subequations}\begin{eqnarray}
v_1(x,y)  &\approx& v^{(0)}_1 + \sigma^p v^{(1)}_{1p}
 + \sigma^p \sigma^q v^{(2)}_{1pq} \\
2H_1 v_1 &=& - K v0
\end{eqnarray}\end{subequations}
\begin{subequations}\begin{eqnarray}
v_2(x,y) &\approx& v^{(0)}_2 \\
4H_2 v_2 &=& - K v1
\end{eqnarray}\end{subequations}
along with
\begin{subequations}\begin{eqnarray}
w_1(x,y) &\approx& w^{(0)}_1 + \sigma^p w^{(1)}_{1p}
 + \sigma^p \sigma^q w^{(2)}_{1pq} \\
2H_1 w_1 + 2H_2 v1 &=& 0
\end{eqnarray}\end{subequations}
\begin{subequations}\begin{eqnarray}
w_2(x,y) &\approx& w^{(0)}_2 \\
4H_2 w_2 + 2H_4 v_2 &=& - K w1
\end{eqnarray}\end{subequations}

Proceeding as in Appendix \ref{apx-VanD}, we find
\begin{subequations}\begin{eqnarray}
%============================== v0 0   ======================================
v^{(0)}_0 &=& 0\\ \cr
%============================== v0 1   ======================================
v^{(1)}_{0a} &=& 0 \\ \cr
%
%============================== v0 2   ======================================
%
v^{(2)}_{0ab} &=&
\left( R{}_{;}{}_{a}{}_{b} - 3\,R{}_{a}{}_{b}{}_{;}{}_{p}{}^{p} +
  4\,R{}_{p}{}_{a}\,R{}_{b}{}^{p} -
  2\,R{}_{p}{}_{q}\,R{}_{a}{}^{q}{}_{b}{}^{p} +
  2\,R{}_{p}{}_{q}{}_{r}{}_{a}\,R{}_{b}{}^{r}{}^{p}{}^{q} \right)/ 360
\\ \cr
%
%============================== v0 3   ======================================
%
v^{(3)}_{0abc} &=& \left(
 5\,R{}_{;}{}_{a}{}_{b}{}_{c} - 14\,R{}_{p}{}_{a}{}_{;}{}_{b}{}_{c}{}^{p} -
  7\,R{}_{a}{}_{b}{}_{;}{}_{p}{}_{c}{}^{p} +
  5\,R{}_{a}{}_{b}{}_{;}{}_{p}{}^{p}{}_{c} +
  8\,R{}_{a}{}_{b}{}_{;}{}_{c}{}_{p}{}^{p} \right. \cr && \hspace{10mm}
+10\,R{}_{p}{}_{a}{}_{;}{}_{b}\,R{}_{c}{}^{p} +
  15\,R{}_{a}{}_{b}{}_{;}{}_{p}\,R{}_{c}{}^{p} -
  10\,R{}_{p}{}_{a}{}_{q}{}_{b}{}_{;}{}_{c}\,R{}^{p}{}^{q} \cr && \hspace{10mm}
 -24\,R{}_{p}{}_{q}{}_{;}{}_{a}\,R{}_{b}{}^{q}{}_{c}{}^{p} +
  4\,R{}_{p}{}_{a}{}_{;}{}_{q}\,R{}_{b}{}^{q}{}_{c}{}^{p} -
  8\,R{}_{p}{}_{a}{}_{q}{}_{r}{}_{;}{}_{b}\,R{}_{c}{}^{p}{}^{q}{}^{r} \cr &&
\hspace{10mm}
+2\,R{}_{p}{}_{a}{}_{q}{}_{b}{}_{;}{}_{r}\,R{}_{c}{}^{p}{}^{q}{}^{r} -
  2\,R{}_{p}{}_{a}{}_{q}{}_{r}{}_{;}{}_{b}\,R{}_{c}{}^{q}{}^{p}{}^{r} +
  2\,R{}_{p}{}_{a}{}_{q}{}_{b}{}_{;}{}_{r}\,R{}_{c}{}^{q}{}^{p}{}^{r} \cr &&
\hspace{10mm} \left.
+2\,R{}_{p}{}_{a}{}_{q}{}_{r}{}_{;}{}_{b}\,R{}_{c}{}^{r}{}^{p}{}^{q} \right) /
    1440
\\ \cr
%
%============================== v0 4   ======================================
%
v^{(4)}_{0abcd} &=& \left(
 -3150\,R{}_{p}{}_{a}{}_{;}{}_{b}\,R{}_{c}{}_{d}{}^{;}{}^{p} -
  2205\,R{}_{a}{}_{b}{}_{;}{}_{p}\,R{}_{c}{}_{d}{}^{;}{}^{p} -
  720\,R{}_{p}{}_{a}{}_{;}{}_{b}\,R{}_{c}{}^{p}{}_{;}{}_{d} \right. \cr &&
+ 11160\,R{}_{p}{}_{q}{}_{;}{}_{a}\,R{}_{b}{}^{p}{}_{c}{}^{q}{}_{;}{}_{d} -
  3600\,R{}_{p}{}_{a}{}_{;}{}_{q}\,R{}_{b}{}^{p}{}_{c}{}^{q}{}_{;}{}_{d} +
  270\,R{}_{p}{}_{a}{}_{q}{}_{b}{}_{;}{}_{r}\,
   R{}_{c}{}^{p}{}_{d}{}^{q}{}^{;}{}^{r} \cr &&
+ 2556\,R{}_{p}{}_{q}{}_{r}{}_{a}{}_{;}{}_{b}\,
   R{}_{c}{}^{p}{}_{d}{}^{r}{}^{;}{}^{q} +
  36\,R{}_{p}{}_{q}{}_{r}{}_{a}{}_{;}{}_{b}\,
   R{}_{c}{}^{p}{}^{q}{}^{r}{}_{;}{}_{d} -
  522\,R{}_{p}{}_{a}{}_{q}{}_{b}{}_{;}{}_{r}\,
   R{}_{c}{}^{p}{}^{q}{}^{r}{}_{;}{}_{d} \cr &&
 -522\,R{}_{p}{}_{a}{}_{q}{}_{b}{}_{;}{}_{r}\,
   R{}_{c}{}^{q}{}^{p}{}^{r}{}_{;}{}_{d} +
  468\,R{}_{p}{}_{q}{}_{r}{}_{a}{}_{;}{}_{b}\,
   R{}_{c}{}^{r}{}^{p}{}^{q}{}_{;}{}_{d} -
  1170\,R{}_{;}{}_{a}{}_{b}{}_{c}{}_{d} \cr &&
+ 1656\,R{}_{p}{}_{a}{}_{;}{}_{b}{}_{c}{}_{d}{}^{p} +
  1404\,R{}_{p}{}_{a}{}_{;}{}_{b}{}_{c}{}^{p}{}_{d} +
  828\,R{}_{a}{}_{b}{}_{;}{}_{p}{}_{c}{}_{d}{}^{p} \cr &&
+ 702\,R{}_{a}{}_{b}{}_{;}{}_{p}{}_{c}{}^{p}{}_{d} -
  810\,R{}_{a}{}_{b}{}_{;}{}_{p}{}^{p}{}_{c}{}_{d} +
  828\,R{}_{a}{}_{b}{}_{;}{}_{c}{}_{p}{}_{d}{}^{p} \cr &&
 -1188\,R{}_{a}{}_{b}{}_{;}{}_{c}{}_{p}{}^{p}{}_{d} -
  1440\,R{}_{a}{}_{b}{}_{;}{}_{c}{}_{d}{}_{p}{}^{p} +
  210\,R{}_{;}{}_{a}{}_{b}\,R{}_{c}{}_{d} \cr &&
 -630\,R{}_{a}{}_{b}{}_{;}{}_{p}{}^{p}\,R{}_{c}{}_{d} -
  3996\,R{}_{p}{}_{a}{}_{;}{}_{b}{}_{c}\,R{}_{d}{}^{p} -
  4914\,R{}_{a}{}_{b}{}_{;}{}_{p}{}_{c}\,R{}_{d}{}^{p} \cr &&
+ 1008\,R{}_{a}{}_{b}{}_{;}{}_{c}{}_{p}\,R{}_{d}{}^{p} +
  840\,R{}_{p}{}_{a}\,R{}_{b}{}_{c}\,R{}_{d}{}^{p} +
  2916\,R{}_{p}{}_{a}{}_{q}{}_{b}{}_{;}{}_{c}{}_{d}\,R{}^{p}{}^{q} \cr &&
 -336\,R{}_{;}{}_{p}{}_{a}\,R{}_{b}{}_{c}{}_{d}{}^{p} +
  1008\,R{}_{p}{}_{a}{}_{;}{}_{q}{}^{q}\,R{}_{b}{}_{c}{}_{d}{}^{p} +
  1344\,R{}_{p}{}_{q}\,R{}_{a}{}^{q}\,R{}_{b}{}_{c}{}_{d}{}^{p} \cr &&
 -336\,R{}_{p}{}_{a}\,R{}_{q}{}_{r}{}_{b}{}^{p}\,R{}_{c}{}_{d}{}^{q}{}^{r} +
  10152\,R{}_{p}{}_{q}{}_{;}{}_{a}{}_{b}\,R{}_{c}{}^{p}{}_{d}{}^{q} +
  2088\,R{}_{p}{}_{a}{}_{;}{}_{q}{}_{b}\,R{}_{c}{}^{p}{}_{d}{}^{q} \cr &&
 -4320\,R{}_{p}{}_{a}{}_{;}{}_{b}{}_{q}\,R{}_{c}{}^{p}{}_{d}{}^{q} -
  1728\,R{}_{a}{}_{b}{}_{;}{}_{p}{}_{q}\,R{}_{c}{}^{p}{}_{d}{}^{q} +
  672\,R{}_{p}{}_{a}\,R{}_{q}{}_{b}\,R{}_{c}{}^{p}{}_{d}{}^{q} \cr &&
 -420\,R{}_{p}{}_{q}\,R{}_{a}{}_{b}\,R{}_{c}{}^{p}{}_{d}{}^{q} +
  4728\,R{}_{p}{}_{a}\,R{}_{q}{}_{b}{}_{r}{}^{p}\,R{}_{c}{}^{q}{}_{d}{}^{r} +
  1536\,R{}_{p}{}_{q}\,R{}_{r}{}_{a}{}_{b}{}^{p}\,R{}_{c}{}^{q}{}_{d}{}^{r}
\cr &&
+ 64\,R{}_{p}{}_{q}{}_{r}{}_{a}\,R{}_{s}{}_{b}{}^{p}{}^{r}\,
   R{}_{c}{}^{q}{}_{d}{}^{s} + 64\,R{}_{p}{}_{q}{}_{r}{}_{a}\,
   R{}_{s}{}^{p}{}_{b}{}^{r}\,R{}_{c}{}^{q}{}_{d}{}^{s} +
  3168\,R{}_{p}{}_{q}{}_{r}{}_{a}\,R{}_{s}{}^{r}{}_{b}{}^{p}\,
   R{}_{c}{}^{q}{}_{d}{}^{s} \cr &&
 -1392\,R{}_{p}{}_{q}{}_{r}{}_{a}\,R{}_{s}{}_{b}{}^{p}{}^{q}\,
   R{}_{c}{}^{r}{}_{d}{}^{s} + 72\,
   R{}_{p}{}_{q}{}_{r}{}_{a}{}_{;}{}_{b}{}_{c}\,R{}_{d}{}^{p}{}^{q}{}^{r} -
  540\,R{}_{p}{}_{a}{}_{q}{}_{b}{}_{;}{}_{r}{}_{c}\,R{}_{d}{}^{p}{}^{q}{}^{r}
\cr &&
 -1296\,R{}_{p}{}_{a}{}_{q}{}_{b}{}_{;}{}_{c}{}_{r}\,
   R{}_{d}{}^{p}{}^{q}{}^{r} - 672\,R{}_{p}{}_{q}\,R{}_{r}{}_{a}{}_{b}{}_{c}\,
   R{}_{d}{}^{p}{}^{q}{}^{r} + 1344\,R{}_{p}{}_{q}{}_{r}{}_{a}\,
   R{}_{s}{}_{b}{}_{c}{}^{q}\,R{}_{d}{}^{p}{}^{r}{}^{s} \cr &&
 -540\,R{}_{p}{}_{a}{}_{q}{}_{b}{}_{;}{}_{r}{}_{c}\,
   R{}_{d}{}^{q}{}^{p}{}^{r} - 1296\,
   R{}_{p}{}_{a}{}_{q}{}_{b}{}_{;}{}_{c}{}_{r}\,R{}_{d}{}^{q}{}^{p}{}^{r} -
  252\,R{}_{p}{}_{a}\,R{}_{q}{}_{b}{}_{r}{}_{c}\,R{}_{d}{}^{q}{}^{p}{}^{r} \cr
&&
+ 432\,R{}_{p}{}_{q}{}_{r}{}_{a}{}_{;}{}_{b}{}_{c}\,R{}_{d}{}^{r}{}^{p}{}^{q} +
  420\,R{}_{a}{}_{b}\,R{}_{p}{}_{q}{}_{r}{}_{c}\,R{}_{d}{}^{r}{}^{p}{}^{q} -
  252\,R{}_{p}{}_{a}\,R{}_{q}{}_{b}{}_{r}{}_{c}\,R{}_{d}{}^{r}{}^{p}{}^{q} \cr
&&
+ 1344\,R{}_{p}{}_{q}{}_{r}{}_{a}\,R{}_{s}{}_{b}{}_{c}{}^{q}\,
   R{}_{d}{}^{r}{}^{p}{}^{s} - 128\,R{}_{p}{}_{q}{}_{r}{}_{a}\,
   R{}_{s}{}_{b}{}_{c}{}^{p}\,R{}_{d}{}^{r}{}^{q}{}^{s} +
  96\,R{}_{p}{}_{q}{}_{r}{}_{a}\,R{}_{s}{}_{b}{}_{c}{}^{r}\,
   R{}_{d}{}^{s}{}^{p}{}^{q} \cr &&
 -128\,R{}_{p}{}_{q}{}_{r}{}_{a}\,R{}_{s}{}_{b}{}_{c}{}^{p}\,
   R{}_{d}{}^{s}{}^{q}{}^{r} - 672\,R{}_{p}{}_{q}{}_{r}{}_{a}\,
   R{}_{s}{}_{b}{}_{c}{}_{d}\,R{}^{p}{}^{q}{}^{r}{}^{s} -
  1200\,R{}_{p}{}_{a}{}_{q}{}_{b}\,R{}_{r}{}_{c}{}_{s}{}_{d}\,
   R{}^{p}{}^{r}{}^{q}{}^{s} \cr && \left.
 -624\,R{}_{p}{}_{a}{}_{q}{}_{b}\,R{}_{r}{}_{c}{}_{s}{}_{d}\,
  R{}^{p}{}^{s}{}^{q}{}^{r} \right)/
907200
\end{eqnarray}\end{subequations}

\begin{subequations}\begin{eqnarray}
%============================== v1 0   ======================================
%
v^{(0)}_1 &=& \left(
R{}_{;}{}_{p}{}^{p} - R{}_{p}{}_{q}\,R{}^{p}{}^{q} +
  R{}_{p}{}_{q}{}_{r}{}_{s}\,R{}^{p}{}^{q}{}^{r}{}^{s}
\right)/ 360
\\ \cr
%
%============================== v1 1   ======================================
%
v^{(1)}_{1a} &=& \left(
-6\,R{}_{;}{}_{p}{}_{a}{}^{p} - 14\,R{}_{;}{}_{p}{}^{p}{}_{a} +
  28\,R{}_{p}{}_{a}{}_{;}{}_{q}{}^{q}{}^{p} - 22\,R{}_{;}{}_{p}\,R{}_{a}{}^{p}
 \right. \cr &&
+40\,R{}_{p}{}_{q}{}_{;}{}_{a}\,R{}^{p}{}^{q} -
  56\,R{}_{p}{}_{a}{}_{;}{}_{q}\,R{}^{p}{}^{q} -
  37\,R{}_{p}{}_{q}{}_{;}{}_{r}\,R{}_{a}{}^{p}{}^{q}{}^{r} -
  19\,R{}_{p}{}_{q}{}_{;}{}_{r}\,R{}_{a}{}^{q}{}^{p}{}^{r} \cr && \left.
 -17\,R{}_{p}{}_{q}{}_{r}{}_{s}{}_{;}{}_{a}\,R{}^{p}{}^{q}{}^{r}{}^{s} +
  2\,R{}_{p}{}_{q}{}_{r}{}_{s}{}_{;}{}_{a}\,R{}^{p}{}^{s}{}^{q}{}^{r} +
  12\,R{}_{p}{}_{a}{}_{q}{}_{r}{}_{;}{}_{s}\,R{}^{p}{}^{s}{}^{q}{}^{r}
\right)/  4320
\\ \cr
%
%============================== v1 2   ======================================
%
v^{(2)}_{1ab} &=& \left(
 210\,R{}_{;}{}_{p}\,R{}_{a}{}_{b}{}^{;}{}^{p} +
  380\,R{}_{;}{}_{p}\,R{}_{a}{}^{p}{}_{;}{}_{b} -
  360\,R{}_{p}{}_{a}{}_{;}{}_{q}\,R{}_{b}{}^{p}{}^{;}{}^{q} -
  2280\,R{}_{p}{}_{a}{}_{;}{}_{q}\,R{}_{b}{}^{q}{}^{;}{}^{p} \right.\cr &&
 -800\,R{}_{p}{}_{q}{}_{;}{}_{a}\,R{}^{p}{}^{q}{}_{;}{}_{b} -
  200\,R{}_{p}{}_{a}{}_{;}{}_{q}\,R{}^{p}{}^{q}{}_{;}{}_{b} -
  1488\,R{}_{;}{}_{p}{}_{a}{}_{b}{}^{p} + 528\,R{}_{;}{}_{p}{}_{a}{}^{p}{}_{b}
\cr &&
+ 820\,R{}_{;}{}_{p}{}^{p}{}_{a}{}_{b} + 180\,R{}_{;}{}_{a}{}_{b}{}_{p}{}^{p} +
  960\,R{}_{p}{}_{q}{}_{;}{}_{a}{}_{b}{}^{p}{}^{q} +
  960\,R{}_{p}{}_{a}{}_{;}{}_{q}{}_{b}{}^{p}{}^{q} \cr &&
+ 960\,R{}_{p}{}_{a}{}_{;}{}_{q}{}_{b}{}^{q}{}^{p} -
  1056\,R{}_{p}{}_{a}{}_{;}{}_{q}{}^{q}{}_{b}{}^{p} -
  944\,R{}_{p}{}_{a}{}_{;}{}_{q}{}^{q}{}^{p}{}_{b} +
  960\,R{}_{p}{}_{a}{}_{;}{}_{b}{}_{q}{}^{p}{}^{q} \cr &&
 -1560\,R{}_{p}{}_{a}{}_{;}{}_{b}{}_{q}{}^{q}{}^{p} +
  480\,R{}_{a}{}_{b}{}_{;}{}_{p}{}_{q}{}^{p}{}^{q} -
  780\,R{}_{a}{}_{b}{}_{;}{}_{p}{}_{q}{}^{q}{}^{p} +
  480\,R{}_{a}{}_{b}{}_{;}{}_{p}{}^{p}{}_{q}{}^{q} \cr &&
+ 140\,R{}_{;}{}_{p}{}^{p}\,R{}_{a}{}_{b} -
  152\,R{}_{;}{}_{p}{}_{a}\,R{}_{b}{}^{p} -
  264\,R{}_{p}{}_{a}{}_{;}{}_{q}{}^{q}\,R{}_{b}{}^{p} -
  1232\,R{}_{p}{}_{q}{}_{;}{}_{a}{}_{b}\,R{}^{p}{}^{q} \cr &&
+ 1768\,R{}_{p}{}_{a}{}_{;}{}_{q}{}_{b}\,R{}^{p}{}^{q} -
  1056\,R{}_{p}{}_{a}{}_{;}{}_{b}{}_{q}\,R{}^{p}{}^{q} -
  132\,R{}_{a}{}_{b}{}_{;}{}_{p}{}_{q}\,R{}^{p}{}^{q} \cr &&
 -3680\,R{}_{p}{}_{a}\,R{}_{q}{}_{b}\,R{}^{p}{}^{q} -
  140\,R{}_{p}{}_{q}\,R{}_{a}{}_{b}\,R{}^{p}{}^{q} \cr &&
+ 1320\,R{}_{p}{}_{q}{}_{;}{}_{r}\,R{}_{a}{}^{p}{}_{b}{}^{q}{}^{;}{}^{r} +
  360\,R{}_{p}{}_{q}{}_{;}{}_{r}\,R{}_{a}{}^{p}{}_{b}{}^{r}{}^{;}{}^{q} +
  5560\,R{}_{p}{}_{q}{}_{;}{}_{r}\,R{}_{a}{}^{p}{}^{q}{}^{r}{}_{;}{}_{b} \cr &&
 -516\,R{}_{p}{}_{q}{}_{r}{}_{a}{}_{;}{}_{s}\,
   R{}_{b}{}^{p}{}^{q}{}^{r}{}^{;}{}^{s} +
  180\,R{}_{p}{}_{q}{}_{r}{}_{a}{}_{;}{}_{s}\,
   R{}_{b}{}^{p}{}^{r}{}^{s}{}^{;}{}^{q} +
  12\,R{}_{p}{}_{q}{}_{r}{}_{a}{}_{;}{}_{s}\,
   R{}_{b}{}^{r}{}^{p}{}^{q}{}^{;}{}^{s} \cr &&
+ 180\,R{}_{p}{}_{q}{}_{r}{}_{a}{}_{;}{}_{s}\,
   R{}_{b}{}^{r}{}^{p}{}^{s}{}^{;}{}^{q} +
  705\,R{}_{p}{}_{q}{}_{r}{}_{s}{}_{;}{}_{a}\,
   R{}^{p}{}^{q}{}^{r}{}^{s}{}_{;}{}_{b} -
  1716\,R{}_{p}{}_{q}{}_{r}{}_{a}{}_{;}{}_{s}\,
   R{}^{p}{}^{q}{}^{r}{}^{s}{}_{;}{}_{b} \cr &&
 -1008\,R{}_{p}{}_{q}{}_{r}{}_{a}{}_{;}{}_{s}\,
   R{}^{p}{}^{r}{}^{q}{}^{s}{}_{;}{}_{b} +
  544\,R{}_{;}{}_{p}{}_{q}\,R{}_{a}{}^{p}{}_{b}{}^{q} -
  72\,R{}_{p}{}_{q}{}_{;}{}_{r}{}^{r}\,R{}_{a}{}^{p}{}_{b}{}^{q} \cr &&
+ 1256\,R{}_{p}{}_{q}\,R{}_{r}{}^{p}\,R{}_{a}{}^{q}{}_{b}{}^{r} +
  6832\,R{}_{p}{}_{q}{}_{;}{}_{r}{}_{a}\,R{}_{b}{}^{p}{}^{q}{}^{r} -
  672\,R{}_{p}{}_{q}{}_{;}{}_{a}{}_{r}\,R{}_{b}{}^{p}{}^{q}{}^{r} \cr &&
+ 1728\,R{}_{p}{}_{a}{}_{;}{}_{q}{}_{r}\,R{}_{b}{}^{p}{}^{q}{}^{r} -
  4712\,R{}_{p}{}_{q}\,R{}_{r}{}_{a}\,R{}_{b}{}^{p}{}^{q}{}^{r} +
  720\,R{}_{p}{}_{a}{}_{;}{}_{q}{}_{r}\,R{}_{b}{}^{q}{}^{p}{}^{r} \cr &&
 -246\,R{}_{p}{}_{q}\,R{}_{r}{}_{s}{}_{a}{}^{p}\,R{}_{b}{}^{q}{}^{r}{}^{s} -
  784\,R{}_{p}{}_{q}\,R{}_{r}{}_{a}{}_{s}{}^{p}\,R{}_{b}{}^{q}{}^{r}{}^{s} -
  276\,R{}_{p}{}_{q}\,R{}_{r}{}_{s}{}_{a}{}^{p}\,R{}_{b}{}^{r}{}^{q}{}^{s} \cr
&&
 -464\,R{}_{p}{}_{q}\,R{}_{r}{}_{a}{}_{s}{}^{p}\,R{}_{b}{}^{r}{}^{q}{}^{s} +
  1472\,R{}_{p}{}_{q}\,R{}_{r}{}_{a}{}_{s}{}^{p}\,R{}_{b}{}^{s}{}^{q}{}^{r} +
  664\,R{}_{p}{}_{q}{}_{r}{}_{s}{}_{;}{}_{a}{}_{b}\,R{}^{p}{}^{q}{}^{r}{}^{s}
\cr &&
 -576\,R{}_{p}{}_{q}{}_{r}{}_{a}{}_{;}{}_{s}{}_{b}\,
   R{}^{p}{}^{q}{}^{r}{}^{s} - 1416\,
   R{}_{p}{}_{q}{}_{r}{}_{a}{}_{;}{}_{b}{}_{s}\,R{}^{p}{}^{q}{}^{r}{}^{s} +
  140\,R{}_{a}{}_{b}\,R{}_{p}{}_{q}{}_{r}{}_{s}\,R{}^{p}{}^{q}{}^{r}{}^{s} \cr
&&
 -1020\,R{}_{p}{}_{q}{}_{r}{}_{a}\,R{}_{s}{}_{t}{}_{b}{}^{r}\,
   R{}^{p}{}^{q}{}^{s}{}^{t} + 64\,
   R{}_{p}{}_{q}{}_{r}{}_{s}{}_{;}{}_{a}{}_{b}\,R{}^{p}{}^{r}{}^{q}{}^{s} -
  96\,R{}_{p}{}_{q}{}_{r}{}_{a}{}_{;}{}_{s}{}_{b}\,R{}^{p}{}^{r}{}^{q}{}^{s}
\cr &&
 -432\,R{}_{p}{}_{q}{}_{r}{}_{a}{}_{;}{}_{b}{}_{s}\,
   R{}^{p}{}^{r}{}^{q}{}^{s} - 192\,
   R{}_{p}{}_{a}{}_{q}{}_{b}{}_{;}{}_{r}{}_{s}\,R{}^{p}{}^{r}{}^{q}{}^{s} -
  2576\,R{}_{p}{}_{q}\,R{}_{r}{}_{a}{}_{s}{}_{b}\,R{}^{p}{}^{r}{}^{q}{}^{s}
\cr &&
 -432\,R{}_{p}{}_{q}{}_{r}{}_{a}\,R{}_{s}{}_{t}{}_{b}{}^{q}\,
   R{}^{p}{}^{r}{}^{s}{}^{t} + 1624\,R{}_{p}{}_{a}\,
   R{}_{q}{}_{r}{}_{s}{}_{b}\,R{}^{p}{}^{s}{}^{q}{}^{r} +
  240\,R{}_{p}{}_{q}{}_{r}{}_{a}\,R{}_{s}{}_{t}{}_{b}{}^{r}\,
   R{}^{p}{}^{s}{}^{q}{}^{t} \cr &&
+ 240\,R{}_{p}{}_{q}{}_{r}{}_{a}\,R{}_{s}{}_{t}{}_{b}{}^{q}\,
   R{}^{p}{}^{s}{}^{r}{}^{t} + 992\,R{}_{p}{}_{q}{}_{r}{}_{s}\,
   R{}_{t}{}_{a}{}_{b}{}^{q}\,R{}^{p}{}^{t}{}^{r}{}^{s} -
  16\,R{}_{p}{}_{q}{}_{r}{}_{a}\,R{}_{s}{}_{t}{}_{b}{}^{p}\,
   R{}^{q}{}^{r}{}^{s}{}^{t} \cr &&
 -2068\,R{}_{p}{}_{q}{}_{r}{}_{a}\,R{}_{s}{}_{b}{}_{t}{}^{p}\,
   R{}^{q}{}^{s}{}^{r}{}^{t} - 1972\,R{}_{p}{}_{q}{}_{r}{}_{a}\,
   R{}_{s}{}_{t}{}_{b}{}^{p}\,R{}^{q}{}^{t}{}^{r}{}^{s} \cr && \left.
 -2988\,R{}_{p}{}_{q}{}_{r}{}_{a}\,R{}_{s}{}_{b}{}_{t}{}^{p}\,
  R{}^{q}{}^{t}{}^{r}{}^{s}
\right)/ 604800
\end{eqnarray}\end{subequations}

\begin{eqnarray}
%============================== v2 0   ======================================
%
v^{(0)}_2 &=& \left(
 139\,R{}_{;}{}_{p}\,R{}^{;}{}^{p} -
  354\,R{}_{p}{}_{q}{}_{;}{}_{r}\,R{}^{p}{}^{q}{}^{;}{}^{r} +
  732\,R{}_{p}{}_{q}{}_{;}{}_{r}\,R{}^{p}{}^{r}{}^{;}{}^{q} +
  114\,R{}_{;}{}_{p}{}^{p}{}_{q}{}^{q} \right. \cr &&
 -288\,R{}_{p}{}_{q}{}_{;}{}_{r}{}^{r}{}^{q}{}^{p} +
  508\,R{}_{;}{}_{p}{}_{q}\,R{}^{p}{}^{q} -
  84\,R{}_{p}{}_{q}{}_{;}{}_{r}{}^{r}\,R{}^{p}{}^{q} \cr && \left.
+484\,R{}_{p}{}_{q}\,R{}_{r}{}^{p}\,R{}^{q}{}^{r} \right)/ 302400 \cr &&
+\left(
 405\,R{}_{p}{}_{q}{}_{r}{}_{s}{}_{;}{}_{t}\,
   R{}^{p}{}^{q}{}^{r}{}^{s}{}^{;}{}^{t} +
  810\,R{}_{p}{}_{q}{}_{r}{}_{s}{}_{;}{}_{t}\,
   R{}^{p}{}^{q}{}^{r}{}^{t}{}^{;}{}^{s} +
  288\,R{}_{p}{}_{q}{}_{;}{}_{r}{}_{s}\,R{}^{p}{}^{r}{}^{q}{}^{s} \right. \cr &&
 -5520\,R{}_{p}{}_{q}\,R{}_{r}{}_{s}\,R{}^{p}{}^{r}{}^{q}{}^{s} -
  2352\,R{}_{p}{}_{q}\,R{}_{r}{}_{s}{}_{t}{}^{p}\,R{}^{q}{}^{t}{}^{r}{}^{s}
\cr && \left.
 -3520\,R{}_{p}{}_{q}{}_{r}{}_{s}\,R{}_{t}{}^{p}{}_{u}{}^{r}\,
   R{}^{q}{}^{t}{}^{s}{}^{u} - 640\,R{}_{p}{}_{q}{}_{r}{}_{s}\,
   R{}_{t}{}_{u}{}^{p}{}^{q}\,R{}^{r}{}^{s}{}^{t}{}^{u} \right)/ 3628800
\end{eqnarray}

\begin{subequations}\begin{eqnarray}
%============================== w1 0   ======================================
%
w^{(0)}_1 &=& -\left(
R{}_{;}{}_{p}{}^{p} - R{}_{p}{}_{q}\,R{}^{p}{}^{q} +
  R{}_{p}{}_{q}{}_{r}{}_{s}\,R{}^{p}{}^{q}{}^{r}{}^{s} \right) / 240
\\ \cr
%
%============================== w1 1   ======================================
%
w^{(1)}_{1a} &=& \left(
12\,R{}_{;}{}_{p}{}_{a}{}^{p} + 31\,R{}_{;}{}_{p}{}^{p}{}_{a} -
  56\,R{}_{p}{}_{a}{}_{;}{}_{q}{}^{q}{}^{p} + 44\,R{}_{;}{}_{p}\,R{}_{a}{}^{p}
 \right. \cr &&
-86\,R{}_{p}{}_{q}{}_{;}{}_{a}\,R{}^{p}{}^{q} +
  112\,R{}_{p}{}_{a}{}_{;}{}_{q}\,R{}^{p}{}^{q} +
  74\,R{}_{p}{}_{q}{}_{;}{}_{r}\,R{}_{a}{}^{p}{}^{q}{}^{r} +
  38\,R{}_{p}{}_{q}{}_{;}{}_{r}\,R{}_{a}{}^{q}{}^{p}{}^{r}  \cr && \left.
+40\,R{}_{p}{}_{q}{}_{r}{}_{s}{}_{;}{}_{a}\,R{}^{p}{}^{q}{}^{r}{}^{s} -
  4\,R{}_{p}{}_{q}{}_{r}{}_{s}{}_{;}{}_{a}\,R{}^{p}{}^{s}{}^{q}{}^{r} -
  24\,R{}_{p}{}_{a}{}_{q}{}_{r}{}_{;}{}_{s}\,R{}^{p}{}^{s}{}^{q}{}^{r}
\right) / 6480
\\ \cr
%
%============================== w1 2   ======================================
%
w^{(2)}_{1ab} &=& \left(
 -630\,R{}_{;}{}_{p}\,R{}_{a}{}_{b}{}^{;}{}^{p} -
  1756\,R{}_{;}{}_{p}\,R{}_{a}{}^{p}{}_{;}{}_{b} +
  1080\,R{}_{p}{}_{a}{}_{;}{}_{q}\,R{}_{b}{}^{p}{}^{;}{}^{q} +
  6840\,R{}_{p}{}_{a}{}_{;}{}_{q}\,R{}_{b}{}^{q}{}^{;}{}^{p} \right. \cr &&
+3856\,R{}_{p}{}_{q}{}_{;}{}_{a}\,R{}^{p}{}^{q}{}_{;}{}_{b} -
  968\,R{}_{p}{}_{a}{}_{;}{}_{q}\,R{}^{p}{}^{q}{}_{;}{}_{b} +
  4464\,R{}_{;}{}_{p}{}_{a}{}_{b}{}^{p} -
  1752\,R{}_{;}{}_{p}{}_{a}{}^{p}{}_{b} \cr &&
 -3020\,R{}_{;}{}_{p}{}^{p}{}_{a}{}_{b} -
  540\,R{}_{;}{}_{a}{}_{b}{}_{p}{}^{p} -
  2880\,R{}_{p}{}_{q}{}_{;}{}_{a}{}_{b}{}^{p}{}^{q} -
  2880\,R{}_{p}{}_{a}{}_{;}{}_{q}{}_{b}{}^{p}{}^{q} \cr &&
 -2880\,R{}_{p}{}_{a}{}_{;}{}_{q}{}_{b}{}^{q}{}^{p} +
  3168\,R{}_{p}{}_{a}{}_{;}{}_{q}{}^{q}{}_{b}{}^{p} +
  3616\,R{}_{p}{}_{a}{}_{;}{}_{q}{}^{q}{}^{p}{}_{b} -
  2880\,R{}_{p}{}_{a}{}_{;}{}_{b}{}_{q}{}^{p}{}^{q} \cr &&
+4680\,R{}_{p}{}_{a}{}_{;}{}_{b}{}_{q}{}^{q}{}^{p} -
  1440\,R{}_{a}{}_{b}{}_{;}{}_{p}{}_{q}{}^{p}{}^{q} +
  2340\,R{}_{a}{}_{b}{}_{;}{}_{p}{}_{q}{}^{q}{}^{p} -
  1440\,R{}_{a}{}_{b}{}_{;}{}_{p}{}^{p}{}_{q}{}^{q} \cr &&
 -504\,R{}_{;}{}_{p}{}^{p}\,R{}_{a}{}_{b} -
  160\,R{}_{;}{}_{p}{}_{a}\,R{}_{b}{}^{p} +
  792\,R{}_{p}{}_{a}{}_{;}{}_{q}{}^{q}\,R{}_{b}{}^{p} +
  5152\,R{}_{p}{}_{q}{}_{;}{}_{a}{}_{b}\,R{}^{p}{}^{q} \cr &&
 -6872\,R{}_{p}{}_{a}{}_{;}{}_{q}{}_{b}\,R{}^{p}{}^{q} +
  3168\,R{}_{p}{}_{a}{}_{;}{}_{b}{}_{q}\,R{}^{p}{}^{q} +
  396\,R{}_{a}{}_{b}{}_{;}{}_{p}{}_{q}\,R{}^{p}{}^{q} \cr &&
+11040\,R{}_{p}{}_{a}\,R{}_{q}{}_{b}\,R{}^{p}{}^{q} +
  504\,R{}_{p}{}_{q}\,R{}_{a}{}_{b}\,R{}^{p}{}^{q} \cr &&
 -3960\,R{}_{p}{}_{q}{}_{;}{}_{r}\,R{}_{a}{}^{p}{}_{b}{}^{q}{}^{;}{}^{r} -
  1080\,R{}_{p}{}_{q}{}_{;}{}_{r}\,R{}_{a}{}^{p}{}_{b}{}^{r}{}^{;}{}^{q} -
  18248\,R{}_{p}{}_{q}{}_{;}{}_{r}\,R{}_{a}{}^{p}{}^{q}{}^{r}{}_{;}{}_{b} \cr &&
+1548\,R{}_{p}{}_{q}{}_{r}{}_{a}{}_{;}{}_{s}\,
   R{}_{b}{}^{p}{}^{q}{}^{r}{}^{;}{}^{s} -
  540\,R{}_{p}{}_{q}{}_{r}{}_{a}{}_{;}{}_{s}\,
   R{}_{b}{}^{p}{}^{r}{}^{s}{}^{;}{}^{q} -
  36\,R{}_{p}{}_{q}{}_{r}{}_{a}{}_{;}{}_{s}\,
   R{}_{b}{}^{r}{}^{p}{}^{q}{}^{;}{}^{s} \cr &&
 -540\,R{}_{p}{}_{q}{}_{r}{}_{a}{}_{;}{}_{s}\,
   R{}_{b}{}^{r}{}^{p}{}^{s}{}^{;}{}^{q} -
  2955\,R{}_{p}{}_{q}{}_{r}{}_{s}{}_{;}{}_{a}\,
   R{}^{p}{}^{q}{}^{r}{}^{s}{}_{;}{}_{b} +
  5484\,R{}_{p}{}_{q}{}_{r}{}_{a}{}_{;}{}_{s}\,
   R{}^{p}{}^{q}{}^{r}{}^{s}{}_{;}{}_{b} \cr &&
+3024\,R{}_{p}{}_{q}{}_{r}{}_{a}{}_{;}{}_{s}\,
   R{}^{p}{}^{r}{}^{q}{}^{s}{}_{;}{}_{b} -
  1632\,R{}_{;}{}_{p}{}_{q}\,R{}_{a}{}^{p}{}_{b}{}^{q} +
  216\,R{}_{p}{}_{q}{}_{;}{}_{r}{}^{r}\,R{}_{a}{}^{p}{}_{b}{}^{q} \cr &&
 -3768\,R{}_{p}{}_{q}\,R{}_{r}{}^{p}\,R{}_{a}{}^{q}{}_{b}{}^{r} -
  22064\,R{}_{p}{}_{q}{}_{;}{}_{r}{}_{a}\,R{}_{b}{}^{p}{}^{q}{}^{r} +
  2016\,R{}_{p}{}_{q}{}_{;}{}_{a}{}_{r}\,R{}_{b}{}^{p}{}^{q}{}^{r} \cr &&
 -5184\,R{}_{p}{}_{a}{}_{;}{}_{q}{}_{r}\,R{}_{b}{}^{p}{}^{q}{}^{r} +
  14136\,R{}_{p}{}_{q}\,R{}_{r}{}_{a}\,R{}_{b}{}^{p}{}^{q}{}^{r} -
  2160\,R{}_{p}{}_{a}{}_{;}{}_{q}{}_{r}\,R{}_{b}{}^{q}{}^{p}{}^{r} \cr &&
+738\,R{}_{p}{}_{q}\,R{}_{r}{}_{s}{}_{a}{}^{p}\,R{}_{b}{}^{q}{}^{r}{}^{s} +
  2352\,R{}_{p}{}_{q}\,R{}_{r}{}_{a}{}_{s}{}^{p}\,R{}_{b}{}^{q}{}^{r}{}^{s} +
  828\,R{}_{p}{}_{q}\,R{}_{r}{}_{s}{}_{a}{}^{p}\,R{}_{b}{}^{r}{}^{q}{}^{s} \cr
&&
+1392\,R{}_{p}{}_{q}\,R{}_{r}{}_{a}{}_{s}{}^{p}\,R{}_{b}{}^{r}{}^{q}{}^{s} -
  4416\,R{}_{p}{}_{q}\,R{}_{r}{}_{a}{}_{s}{}^{p}\,R{}_{b}{}^{s}{}^{q}{}^{r} -
  2804\,R{}_{p}{}_{q}{}_{r}{}_{s}{}_{;}{}_{a}{}_{b}\,R{}^{p}{}^{q}{}^{r}{}^{s}
\cr &&
+2064\,R{}_{p}{}_{q}{}_{r}{}_{a}{}_{;}{}_{s}{}_{b}\,
   R{}^{p}{}^{q}{}^{r}{}^{s} + 4248\,
   R{}_{p}{}_{q}{}_{r}{}_{a}{}_{;}{}_{b}{}_{s}\,R{}^{p}{}^{q}{}^{r}{}^{s} -
  504\,R{}_{a}{}_{b}\,R{}_{p}{}_{q}{}_{r}{}_{s}\,R{}^{p}{}^{q}{}^{r}{}^{s} \cr
&&
+3060\,R{}_{p}{}_{q}{}_{r}{}_{a}\,R{}_{s}{}_{t}{}_{b}{}^{r}\,
   R{}^{p}{}^{q}{}^{s}{}^{t} - 248\,
   R{}_{p}{}_{q}{}_{r}{}_{s}{}_{;}{}_{a}{}_{b}\,R{}^{p}{}^{r}{}^{q}{}^{s} +
  288\,R{}_{p}{}_{q}{}_{r}{}_{a}{}_{;}{}_{s}{}_{b}\,R{}^{p}{}^{r}{}^{q}{}^{s}
\cr &&
+1296\,R{}_{p}{}_{q}{}_{r}{}_{a}{}_{;}{}_{b}{}_{s}\,
   R{}^{p}{}^{r}{}^{q}{}^{s} + 576\,
   R{}_{p}{}_{a}{}_{q}{}_{b}{}_{;}{}_{r}{}_{s}\,R{}^{p}{}^{r}{}^{q}{}^{s} +
  7728\,R{}_{p}{}_{q}\,R{}_{r}{}_{a}{}_{s}{}_{b}\,R{}^{p}{}^{r}{}^{q}{}^{s}
\cr &&
+1296\,R{}_{p}{}_{q}{}_{r}{}_{a}\,R{}_{s}{}_{t}{}_{b}{}^{q}\,
   R{}^{p}{}^{r}{}^{s}{}^{t} - 4872\,R{}_{p}{}_{a}\,
   R{}_{q}{}_{r}{}_{s}{}_{b}\,R{}^{p}{}^{s}{}^{q}{}^{r} -
  720\,R{}_{p}{}_{q}{}_{r}{}_{a}\,R{}_{s}{}_{t}{}_{b}{}^{r}\,
   R{}^{p}{}^{s}{}^{q}{}^{t} \cr &&
 -720\,R{}_{p}{}_{q}{}_{r}{}_{a}\,R{}_{s}{}_{t}{}_{b}{}^{q}\,
   R{}^{p}{}^{s}{}^{r}{}^{t} - 2976\,R{}_{p}{}_{q}{}_{r}{}_{s}\,
   R{}_{t}{}_{a}{}_{b}{}^{q}\,R{}^{p}{}^{t}{}^{r}{}^{s} +
  48\,R{}_{p}{}_{q}{}_{r}{}_{a}\,R{}_{s}{}_{t}{}_{b}{}^{p}\,
   R{}^{q}{}^{r}{}^{s}{}^{t} \cr &&
+6204\,R{}_{p}{}_{q}{}_{r}{}_{a}\,R{}_{s}{}_{b}{}_{t}{}^{p}\,
   R{}^{q}{}^{s}{}^{r}{}^{t} + 5916\,R{}_{p}{}_{q}{}_{r}{}_{a}\,
   R{}_{s}{}_{t}{}_{b}{}^{p}\,R{}^{q}{}^{t}{}^{r}{}^{s} \cr && \left.
+8964\,R{}_{p}{}_{q}{}_{r}{}_{a}\,R{}_{s}{}_{b}{}_{t}{}^{p}\,
  R{}^{q}{}^{t}{}^{r}{}^{s} \right)/
1451520
\end{eqnarray}\end{subequations}

\begin{eqnarray}
%============================== w2 0   ======================================
%
w^{(0)}_2 &=& -\left(
 139\,R{}_{;}{}_{p}\,R{}^{;}{}^{p} -
  354\,R{}_{p}{}_{q}{}_{;}{}_{r}\,R{}^{p}{}^{q}{}^{;}{}^{r} +
  732\,R{}_{p}{}_{q}{}_{;}{}_{r}\,R{}^{p}{}^{r}{}^{;}{}^{q} \right. \cr &&
+114\,R{}_{;}{}_{p}{}^{p}{}_{q}{}^{q} -
  288\,R{}_{p}{}_{q}{}_{;}{}_{r}{}^{r}{}^{q}{}^{p} +
  508\,R{}_{;}{}_{p}{}_{q}\,R{}^{p}{}^{q} \cr && \left.
 -84\,R{}_{p}{}_{q}{}_{;}{}_{r}{}^{r}\,R{}^{p}{}^{q} +
  484\,R{}_{p}{}_{q}\,R{}_{r}{}^{p}\,R{}^{q}{}^{r} \right)/
145152 \cr
&&-\left(
 405\,R{}_{p}{}_{q}{}_{r}{}_{s}{}_{;}{}_{t}\,
   R{}^{p}{}^{q}{}^{r}{}^{s}{}^{;}{}^{t} +
  810\,R{}_{p}{}_{q}{}_{r}{}_{s}{}_{;}{}_{t}\,
   R{}^{p}{}^{q}{}^{r}{}^{t}{}^{;}{}^{s} +
  288\,R{}_{p}{}_{q}{}_{;}{}_{r}{}_{s}\,R{}^{p}{}^{r}{}^{q}{}^{s} \right. \cr &&
 -5520\,R{}_{p}{}_{q}\,R{}_{r}{}_{s}\,R{}^{p}{}^{r}{}^{q}{}^{s} -
  2352\,R{}_{p}{}_{q}\,R{}_{r}{}_{s}{}_{t}{}^{p}\,R{}^{q}{}^{t}{}^{r}{}^{s} -
  3520\,R{}_{p}{}_{q}{}_{r}{}_{s}\,R{}_{t}{}^{p}{}_{u}{}^{r}\,
   R{}^{q}{}^{t}{}^{s}{}^{u} \cr && \left.
 -640\,R{}_{p}{}_{q}{}_{r}{}_{s}\,R{}_{t}{}_{u}{}^{p}{}^{q}\,
  R{}^{r}{}^{s}{}^{t}{}^{u} \right)/
1741824
\end{eqnarray}

%%%%%%%%%%%%%%%%%%%%%%%%%%%%%%%%%%%%%%%%%%%%%%%%%%%%%%%
%%%%%%%%%%%%%%%% The Bibliography %%%%%%%%%%%%%%%%%%%%%

\end{document}